\documentclass[a4paper,11pt]{article}
\usepackage{jinstpub} 
\usepackage{lineno}
\usepackage{url}
\usepackage{graphicx}

\usepackage{xcolor}
\usepackage{amsmath}
\usepackage{subcaption}
\usepackage{xspace}
\newcommand{\urwell}{\ensuremath{\mu\mathrm{-RWELL}}\xspace}

\title{\boldmath Performance of resistive MPGDs with pad readout coupled to VMM3a ASIC}

\author[a,b,1]{D. Zavazieva,\note{Corresponding author.}}
\author[b]{I. Ravinovich,}
\author[b]{L. Moleri,}
\author[b]{M. Borysova,}
\author[b]{E. B. Shields,}
\author[b]{O. Arie,}
\author[b]{M. Kwok Lam Chu,}
\author[a]{L. Arazi}
\author[b]{and S. Bressler}

\affiliation[a]{Ben-Gurion University of the Negev,\\David Ben Gurion Blvd 1, Be'er Sheva, Israel}
\affiliation[b]{Weizmann Institute of Science,\\234 Herzl street, Rehovot, Israel}

\emailAdd{darinaza@post.bgu.ac.il}

\abstract{We present a comparative study of three resistive Micro-Pattern Gaseous Detector (MPGD) technologies --- Micromegas, RPWELL, and \urwell ~--- with VMM3a based readout, using relativistic muons and pions. The Micromegas and the RPWELL were operated in $\mathrm{Ar/CO_2/iC_4H_{10}}$ gas mixture, while the \urwell in  $\mathrm{Ar/CO_2/CF_4}$ (greenhouse gas containing mixture). All detectors operated stably exceeding  96\% efficiency. The usage of continuous readout enabled studies of detector electrical instabilities in- and off-beam at near breakdown voltages. Each technology has different advantages making it more suitable for various experimental conditions.}


\arxivnumber{} 

\begin{document}
\maketitle
\flushbottom

\section{Introduction}
\label{sec:intro}

Resistive Micro-Pattern Gaseous Detector (MPGD) technologies — such as the Micro-Mesh Gaseous Structure (Micromegas) \cite{Alviggi:2024lir}, the Resistive Plate WELL (RPWELL) \cite{Rubin:2013jna}, and the Micro Resistive WELL (\urwell) \cite{Bencivenni:2024ryx} — are single-stage amplification detectors in which signals are induced on the readout electrodes through a resistive material. The resistive layer serves as a protective element for the readout electronics, quenching the discharge energy and thereby enhancing operational stability (see, for example,~\cite{jash2024discharge}).

% Past studies reported spatial resolution of X, Y, Z for \urwell, Micromegas and RPWELL, respectively. Time ...

Although each of these technologies has been extensively studied on its own,  not many studies have focused on  comparing them systematically under identical experimental conditions. Exceptions are  a comparison of the Micromegas and the \urwell in the context of a tracking system~\cite{Scharenberg:2024abh}, and a brief comparison of the three technologies in the context of a hadronic calorimeter~\cite{MuonColliderendorsedbytheInternational:2025zcf}. In this work, we present a comprehensive performance comparison of the three detector types, tested with muon and pion beams using a common readout system. The preliminary results of these measurements were reported in \cite{Zavazieva:2025rsj}. Here, we extend the analysis with a more detailed investigation of the detectors’ performance, focusing on efficiency measurement, charge and time response, and operational stability.

\section{Studied prototypes}

The three detector technologies introduced above were implemented on identical readout printed circuit boards (PCB), developed within the framework of a DRD1 collaboration Common Project \cite{Longo:2024ntu}. Each PCB consisted of 384 copper pads measuring an area of $1\times1~\mathrm{cm^2}$, arranged to cover an active area of approximately $20\times20~\mathrm{cm^2}$. The signal routing to the readout electronics was implemented via conductive lines distributed across three inner PCB layers and extracted with three multi-pin connectors\footnote{FX10A-140S14-SV}.

On top of the identical PCBs, three different resistive anodes and three different multiplication electrodes were mounted. The Micromegas detector had an $\approx 100 ~\mu\mathrm{m}$ multiplication gap. A double-layer of diamond-like carbon (DLC) coated Kapton was used as a resistive layer and included a periodic array of conductive vias for charge evacuation to the readout pads \cite{Alviggi:2024lir}. For the \urwell, 50 $\mu$m thick single-sided Gaseous Electron Multiplier (GEM) foil (conical holes, 50 $\mu$m-diameter, 140  $\mu$m-pitch, hexagonal pattern) was glued to a single DLC coated Kapton layer, with grounding lines deposited across the active area at 10~mm intervals \cite{Bencivenni:2024ryx}. The RPWELL detector was composed of a 0.4 mm thick single sided copper clad Thick-Gaseous Electron Multiplier (THGEM) placed against a thick resistive plate. The THGEM holes (0.5 mm-diameter, 0.1 mm-rim, 0.96 mm-pitch) were arranged in a square lattice. Each group of $10\times10$ holes defined a $1\times1~\mathrm{cm^2}$ pad regions separated by a copper grid (1.36 mm-width). The resistive plate made of iron-doped, low bulk-resistivity glass ($2 ~\mathrm{G}\Omega\cdot\mathrm{cm}$)~\cite{WANG2010151} was electrically coupled across the entire area using resistive epoxy-graphite mixture \cite{zavazieva2023towards}.

The signal in the three technologies is induced on the readout pads through the resistive materials. Avalanche charges in both the Micromegas and the RPWELL evacuate vertically through the readout electronics, while in the \urwell ~--- sideways through the ground lines on the surface to a common ground outside of the gas volume. 

All three prototypes were housed in identical gas boxes, composed of a 6 mm thick FR4 frame and a copper-clad FR4 plate acting as the drift electrode. The drift gap for each detector was calculated as $\mathrm{(6 - d_a)}$~mm, where $\mathrm{d_a}$ represents the thickness of the amplification gap for the Micromegas and the \urwell, and the combined thickness of resistive plate and THGEM for the RPWELL. As a result, each detector featured a slightly different drift gap, contributing to variations in the number of primary ionization electrons produced.

Based on prior experience, while the Micromegas and the RPWELL were operated with $\mathrm{Ar/CO_2/iC_4H_{10}}$ (93/5/2) gas mixture, the \urwell was operated with $\mathrm{Ar/CO_2/CF_4}$ (45/15/40) which is environmentally disadvantageous due to the presence of CF$_4$, a potent greenhouse gas. These choices were guided by considerations of detector stability and gain, and also influenced the primary ionization yield due to differing ionization potentials of the gases. A summary of detector-specific parameters relevant to performance evaluation is provided in Table~\ref{tab:summary}.

\begin{table}[!h]
    \centering
    \begin{tabular}{|l|c|c|c|}
        \hline
            & Micromegas &  RPWELL & $\mu$-RWELL \\
        \hline
        Amplification gap [$\mu$m] & $\sim 100$ & $\sim 400$ & $\sim 50$ \\  
        \hline
        Drift gap [mm]   & 5.9 & 4.9 & 5.95  \\
        \hline
        \begin{tabular}{@{}l@{}}Amplification field  \\ (parallel plate approximation) [kV/cm]  \end{tabular}   & 38 – 51 & 26 – 30 &  90 – 118 \\
        \hline
        Working gas composition & \begin{tabular}{@{}c@{}} $\mathrm{Ar/CO_2/iC_4H_{10}}$ \\ 93/5/2 \end{tabular} & \begin{tabular}{@{}c@{}} $\mathrm{Ar/CO_2/iC_4H_{10}}$ \\ 93/5/2 \end{tabular} & \begin{tabular}{@{}c@{}} $\mathrm{Ar/CO_2/CF_4}$ \\ 45/15/40 \end{tabular} \\
        \hline
        Number of primary electrons & 63 & 52  & 65 \\
        \hline
        Resistivity & $\sim 50 ~\mathrm{M}\Omega/\Box$ & $\sim 2 ~\mathrm{G}\Omega\cdot\mathrm{cm}$ & $\sim 100 ~\mathrm{M}\Omega/\Box$ \\
        \hline
        Charge signal rise time [ns] & $\sim 100$ & $\sim 2000$ & $\sim 50$ \\ 
        \hline
    \end{tabular}
    \caption{List of parameters relevant for detector performance.}
    \label{tab:summary}
\end{table}

\section{Experimental setup \& Methodology}
\label{sec:method}

\subsection{Beam tracker and readout system}
\label{sec}

\begin{figure}
    \centering
    \includegraphics[width=0.95\linewidth]{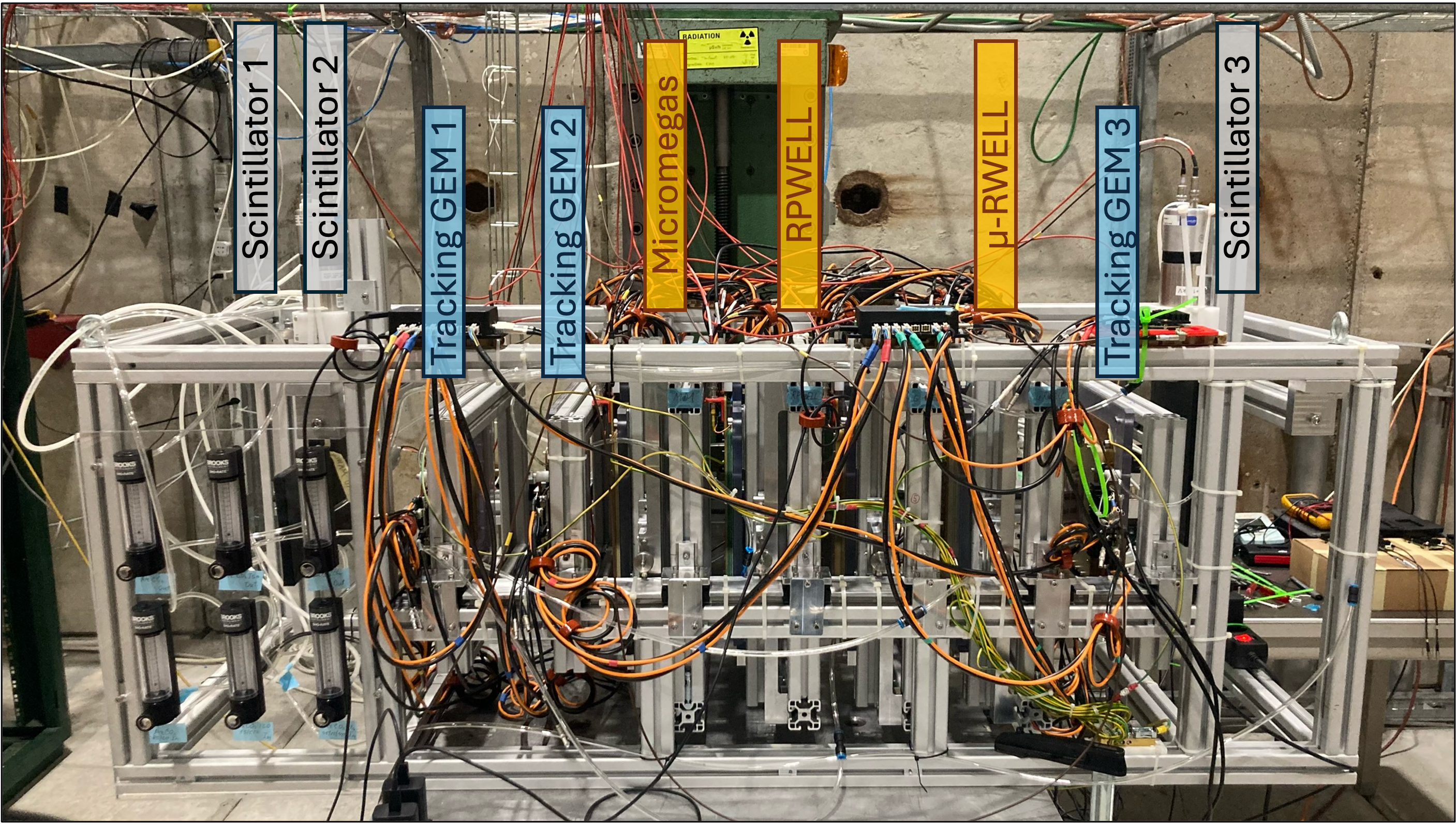}
    \caption{Experimental setup at the H4 beam line, CERN SPS North Area.}
    \label{fig:setup}
\end{figure}

Figure~\ref{fig:setup} shows the experimental setup built for the study. The three tested detectors were integrated into a beam telescope composed of three position-sensitive, COMPASS-like triple-GEM detectors with strip readout \cite{Altunbas:2002ds}, each having an active area of $10\times10~\mathrm{cm^2}$. These tracking detectors were used for particle trajectory reconstruction at a precision\footnote{Resolution measured as the RMS of a Gaussian fit to the spatial residuals in each tracker layer when excluding it from the reconstruction and extrapolating tracks from the other two planes.} of ~40 $\mu\mathrm{m}$. In addition, three  plastic scintillators coupled to  photomultiplier tubes (PMT) were used to provide a sub-ns time reference for the incoming particles. The PMT signals were processed through NIM electronic units for discrimination and coincidence logic. 

The scintillator coincidence, along with the anode signals from the tracking GEMs and tested detectors, was read out using the 10-bit ADC precision  VMM3a-based RD51 Scalable Readout System (SRS) \cite{scharenberg2022development}, comprising a total of 2689 readout channels. Each VMM3a ASIC was operated with a gain of 3 mV/fC and a shaping time of 200~ns. The analog threshold was calibrated individually for each channel: approximately 10~mV for the tested detectors and 30~mV for the tracking detectors. These thresholds correspond to charge values of the order of $(2 - 6)\times10^4$ electrons and were chosen to be roughly ten times the baseline noise level, as observed on the analog VMM3a output using an oscilloscope. As the system operates in continuous (self-triggered) mode, the relatively high threshold values represent a trade-off, aiming to suppress pick-up noise while maintaining good sensitivity to particle signals.

The experimental setup was deployed at the H4 beam line in the North Area of the CERN Super Proton Synchrotron (SPS). The detectors were tested under controlled conditions using both relativistic muon (100 GeV/c) and pion (10 GeV/c) beams.

\subsection{Data processing and analysis}

Signals above the analog threshold were recorded as hits and processed using the \texttt{vmm-sdat} software package \cite{vmmsdat}. Neighboring strips or pads were grouped into a cluster if they were separated by less than 200~ns. For the tracking detectors, one missing strip was tolerated within the cluster. In contrast, no missing pads were allowed for the tested detectors due to their relatively large pad dimensions. 

Although the analog threshold effectively suppressed baseline fluctuations, a high rate of low-amplitude hits was observed in the RPWELL detector which was suppressed with an additional  offline threshold of 50 ADC  applied during the cluster reconstruction step.

\begin{figure}
    \centering
    \includegraphics[width=0.48\linewidth]{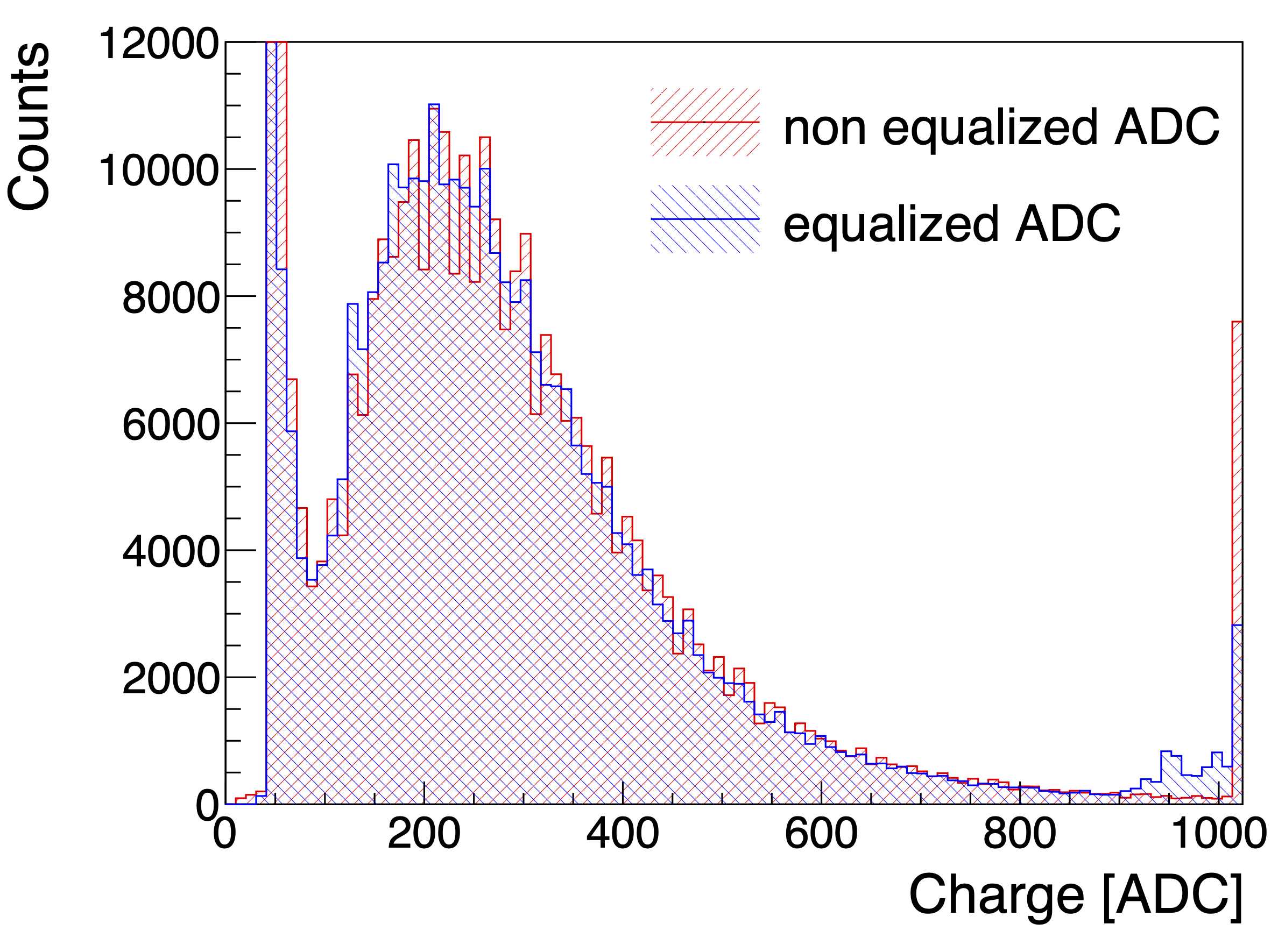}
    \caption{Pad charge measured with the Micromegas detector at 490 V with 0.5 kV/cm.}
    \label{fig:adc_saturation}
\end{figure}

Figure~\ref{fig:adc_saturation} shows the Micromegas pad charge spectra before (red) and after (blue) ADC equalization, which is applied to correct for non-uniformities in the VMM3a charge response across its 64 read-out channels \cite{scharenberg2022development}. At high detector gain, charge saturation occurs, visible as 1023 ADC counts in a channel before the equalization. As a result of the ADC correction, saturated signals originally capped at 1023 ADC counts are redistributed, typically appearing above 900 ADC counts in the calibrated spectrum. Therefore, we define a discharge-like event as a cluster containing at least one pad with charge above 900 ADC. While a fraction of those clusters are likely corresponding to the tail of Landau distribution, conservatively, we count them as discharges.

\begin{figure}
    \centering
    \includegraphics[width=0.95\linewidth]{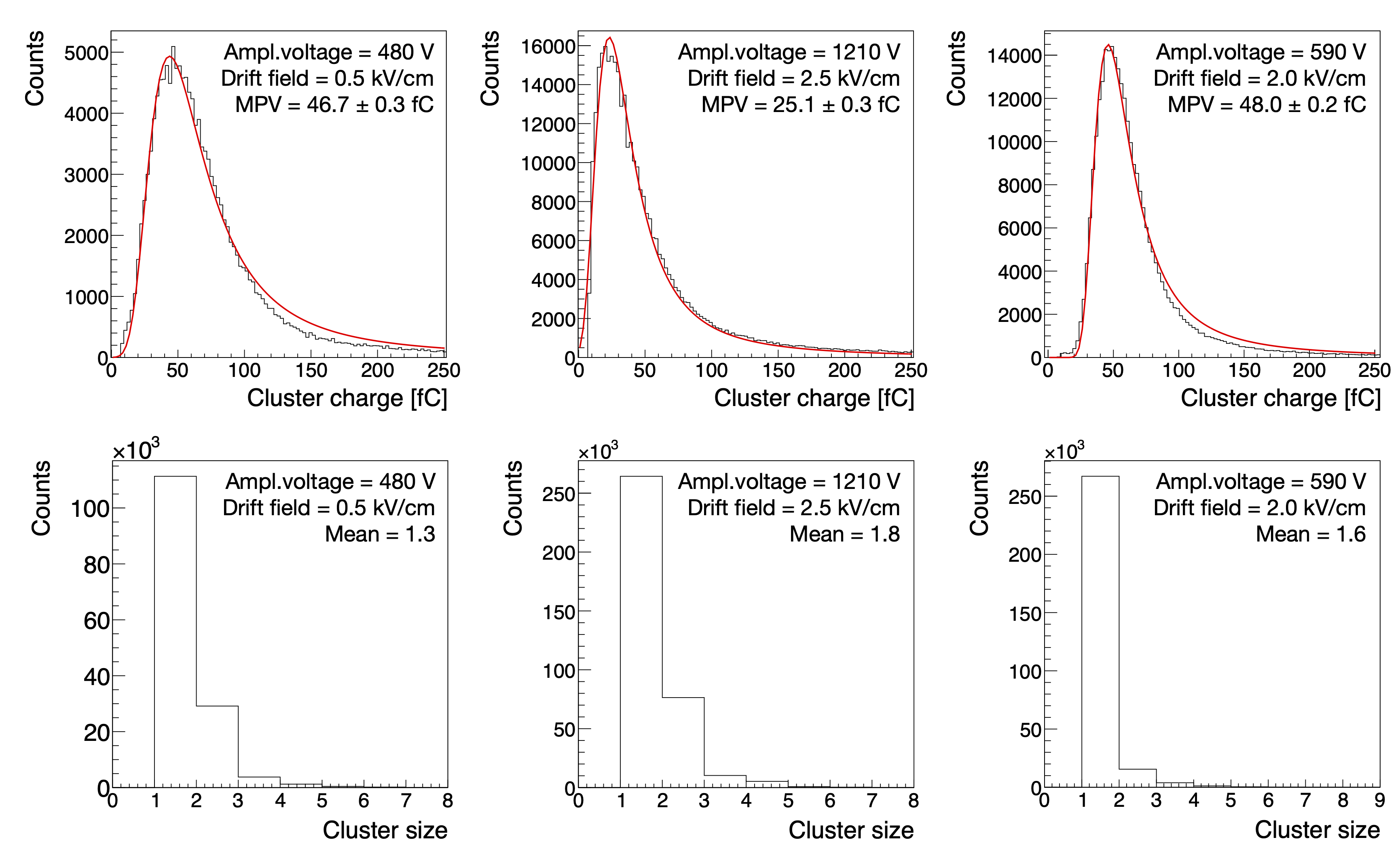}
    \caption{Examples of cluster charge (top) and cluster size (bottom) measured with Micromegas (left), RPWELL (center) and \urwell (right).}
    \label{fig:examples}
\end{figure}

The cluster charge was defined as the sum of all hits' (strips or pads) charges in the cluster, while the cluster position and time were calculated as charge-weighted averages across all hits in the cluster. The measured ADC values were converted to the equivalent charge following a linear calibration function and known test capacitance\footnote{With variation less than 10\% between channels} \cite{scharenberg2022development}. An example of cluster charge and cluster size distributions measured with the three tested detectors is given in Figure \ref{fig:examples} (top row) --- the data was taken at 480, 1210, 590 ~V amplification voltage and 0.5, 2.5, 2.0 ~kV/cm drift field for the Micromegas, RPWELL, \urwell, respectively. While the majority of clusters are of size 1 as shown in Figure ~\ref{fig:examples} (bottom row), larger cluster multiplicity is often measured closer to the readout pads borders where total charge is shared between several pads as reported in \cite{Zavazieva:2025rsj}.

Following cluster reconstruction, an event-building procedure was implemented. Each trigger signal provided time reference for an event which consisted of all clusters measured within $\pm2000$~ns from the trigger time. The time difference was calculated as $\mathrm{dt = t_{cluster} - t_{trigger}}$, where $\mathrm{t_{cluster}}$ is the above mentioned cluster time and $\mathrm{t_{trigger}}$ corresponds to the time of the trigger signal from the scintillator coincidence logic. 

\begin{figure}
    \centering
    \begin{subfigure}[b]{0.48\textwidth} 
    \includegraphics[width=\textwidth]{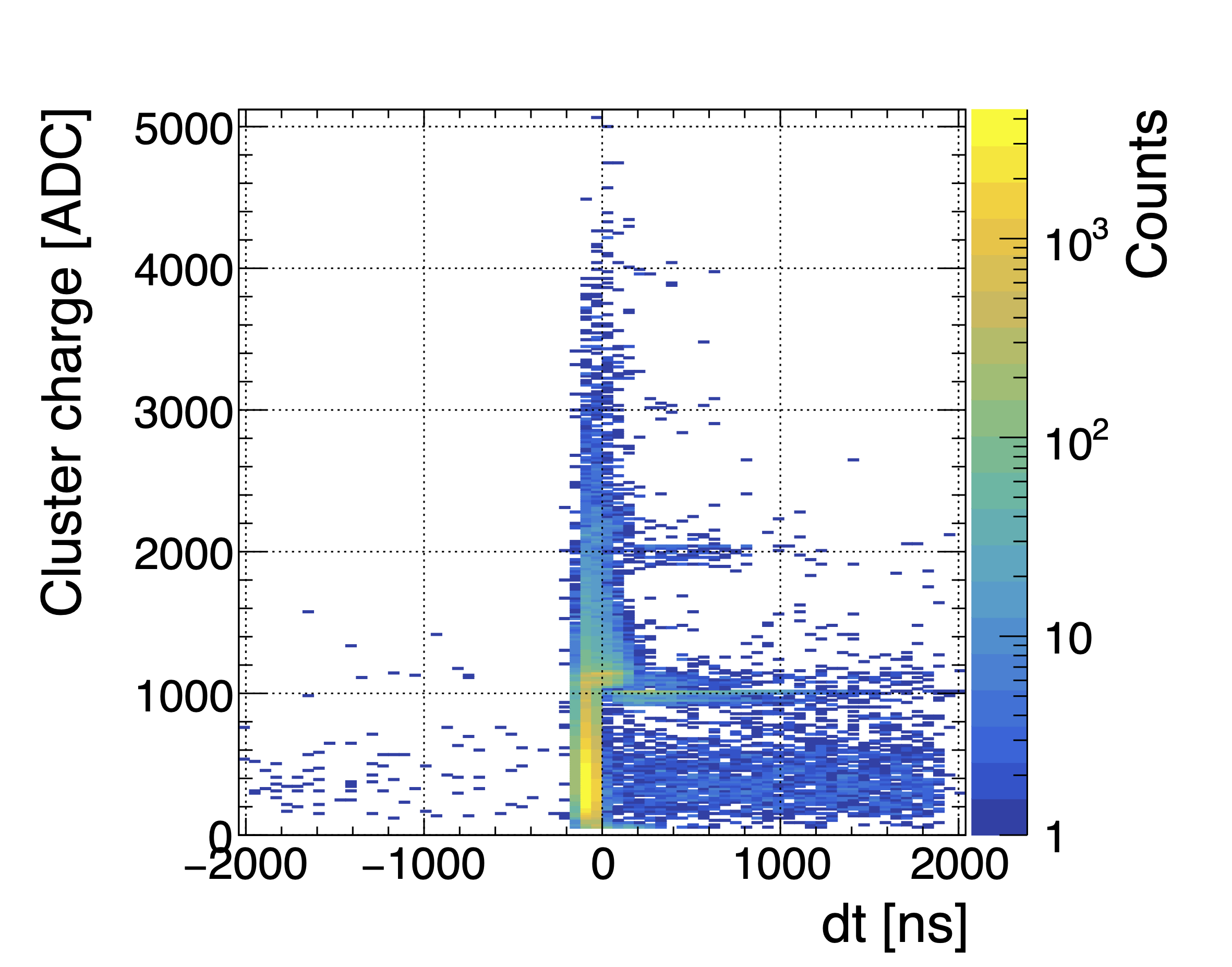}
    \caption{\label{fig:mm_500V_q_vs_dt}Cluster charge vs dt}
    \end{subfigure}
    \begin{subfigure}[b]{0.48\textwidth} 
    \includegraphics[width=\textwidth]{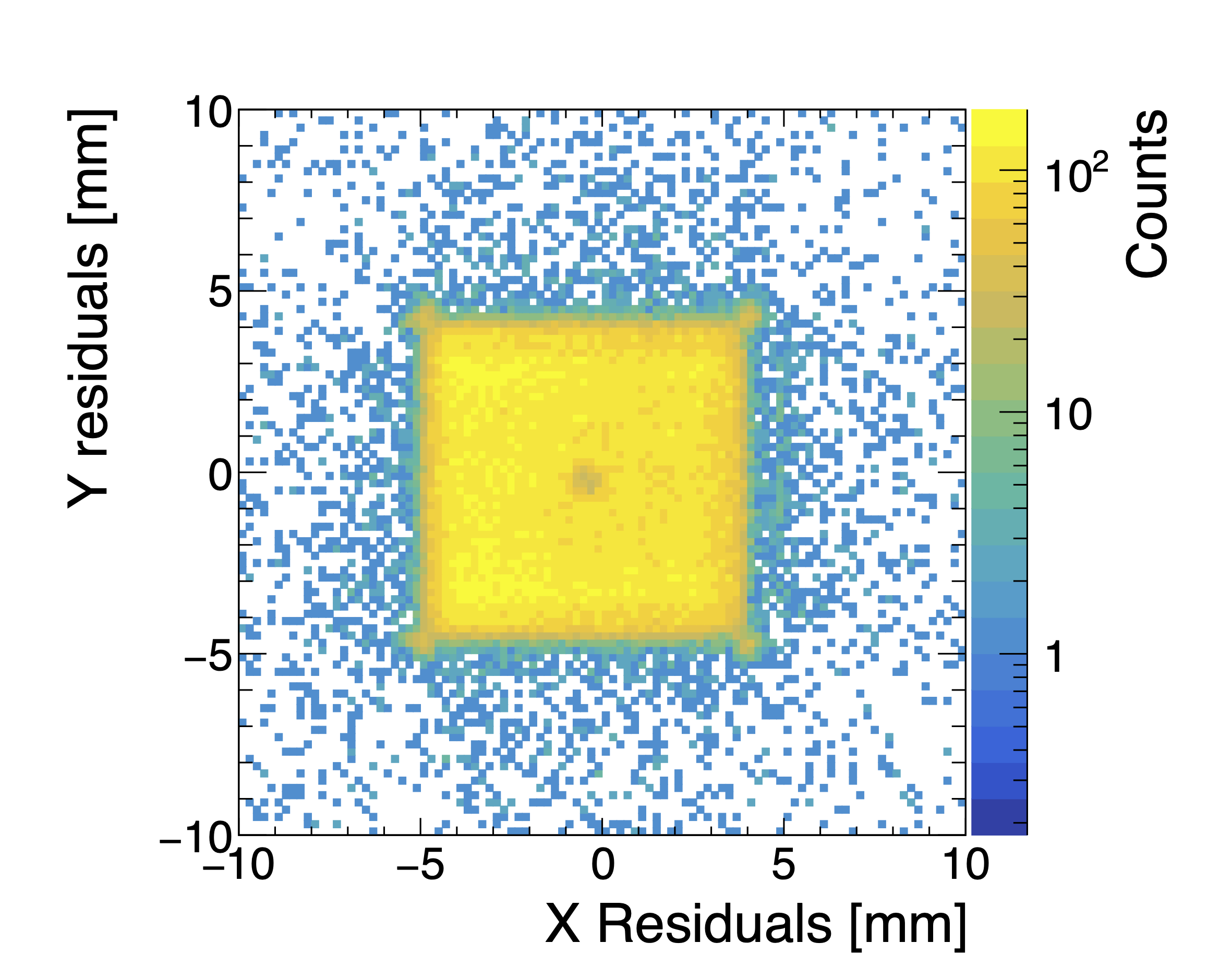}
    \caption{\label{fig:mm_500V_residuals} Residuals}

    \end{subfigure}
    \caption{\label{fig:analysis} Results of the data analysis for Micromegas at 500 V with 0.5 kV/cm: a) dependency of cluster charge on dt, b) spatial residuals in X and Y directions. }
\end{figure}

While the time of muon signals mostly appeared within a few hundred nanoseconds from the trigger, saturated signals were registered with a delay. In Figure \ref{fig:mm_500V_q_vs_dt}, the cluster charge is shown as function of $\mathrm{dt}$. The horizontal lines correspond to clusters with saturated pads, their delay is attributed to the peak-finding function of the VMM3a, returning a time value which could correspond to any point of the large ($1 ~\mu\mathrm{s}$) saturation plateau of the signal. Clusters with charge below 900~ADC and large $\mathrm{dt}$ values are typically uncorrelated with tracks and are excluded during track reconstruction. 

Within each event, the tracker clusters were used for track reconstruction, using either a Hough Transform method — implemented via a custom-developed package \cite{tb24-tracking} — or a straight-line fit made possible with an extended module \cite{corrympgd} of the Corryvreckan framework \cite{dannheim2021corryvreckan}. If all three tracking layers yielded a consistent perpendicular track with slopes constrained to $-0.01 < a_{x,y} < 0.01$ (where $a$ is the track slope), the reconstructed trajectory was extrapolated to the tested detectors' planes. A cluster in the tested detector was considered matched to the track if it was located within $\pm10$~mm of the expected intercept on its plane. This matching window was chosen based on the pad size and the position residuals distribution, as shown in Figure~\ref{fig:mm_500V_residuals}. Matched clusters were included in the efficiency calculation and retained for subsequent analysis. 

Preliminary results reported in \cite{Zavazieva:2025rsj} revealed specific design limitations in the \urwell prototype. In particular, the DLC grounding lines were identified as dead areas, affecting nearly 30\% of the tested region. As future designs aim to replace these lines with grounding dots to significantly reduce inactive regions the affected areas were excluded from the analysis. In the RPWELL prototype, three readout electronic channels were dead during data acquisition and were excluded from the analysis.

The detectors' performance was characterized in terms of gain, efficiency, discharge rate and time resolution. Measurements were conducted by varying the amplification voltage and drift field within ranges appropriate for each technology. Specifically, the amplification voltage was scanned with steps of 10~V between 380 – 510~V for the Micromegas, 1050 – 1210~V for the RPWELL, and 450 – 590~V for the \urwell. Under parallel plate approximation this corresponds to the amplification field ranges of 38 - 51 ~kV/cm for the  Micromegas, 26 - 30~kV/cm for the RPWELL, and 90 – 118~kV/cm for the \urwell. It is important to note that a 10~V increase in amplification voltage corresponds to a much smaller increase in the amplification field in the RPWELL compared to the other two technologies, thus we expect a more gradual change in the RPWELL performance along the scans. The drift field was defined as an electric field applied to the drift gap and varied in a range of 0.5 - 4.5~kV/cm for each tested detector (all estimated using the parallel plate approximation).

\section{Performance}
\label{sec:performance}

\subsection{Efficiency}

\begin{figure}
    \centering
    \begin{subfigure}[b]{0.485\textwidth}
    \includegraphics[width=\textwidth]{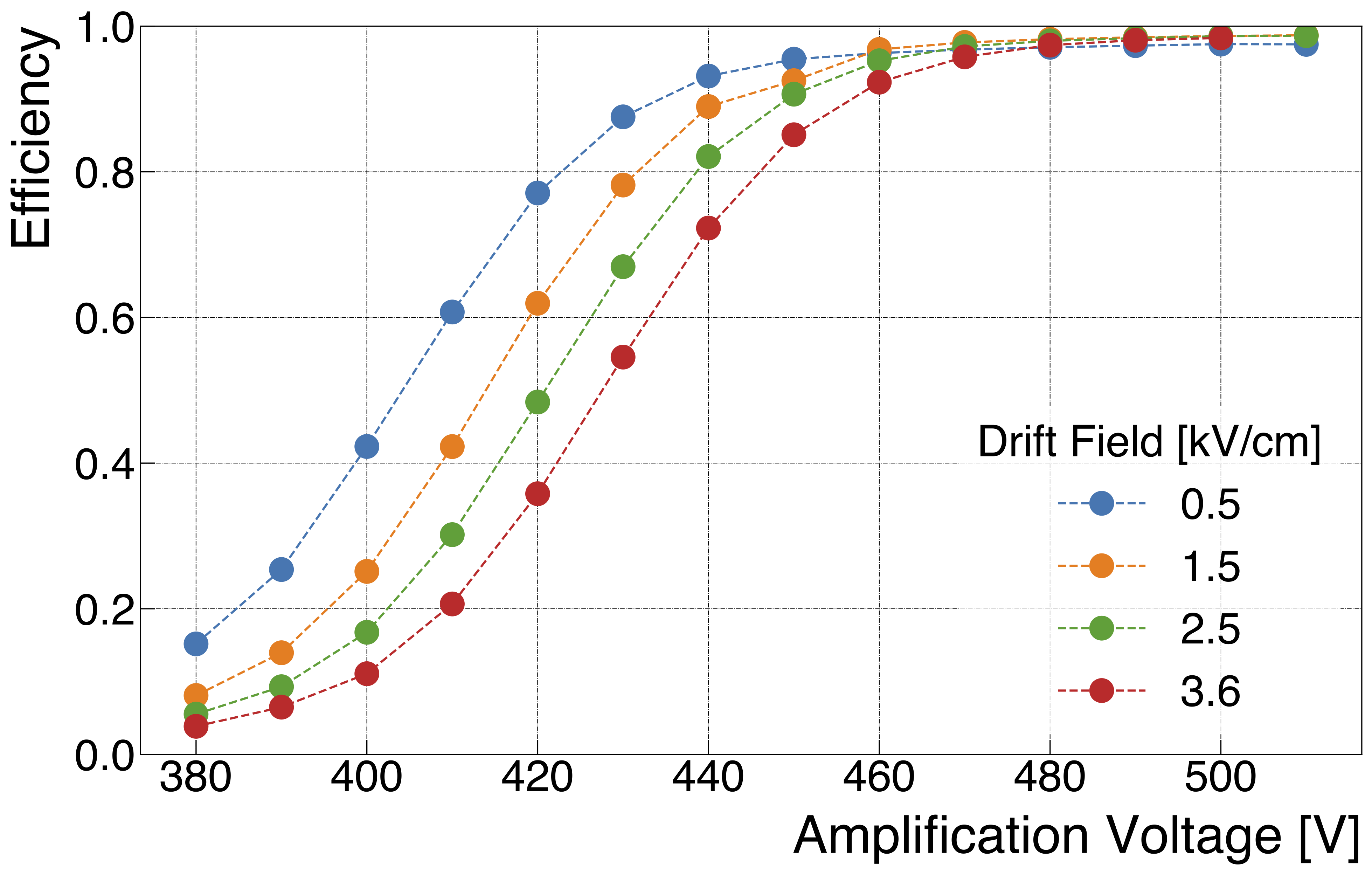}
    \caption{Micromegas}
    \label{fig:efficiency-mm}
    \end{subfigure}
    \begin{subfigure}[b]{0.485\textwidth}
    \includegraphics[width=\textwidth]{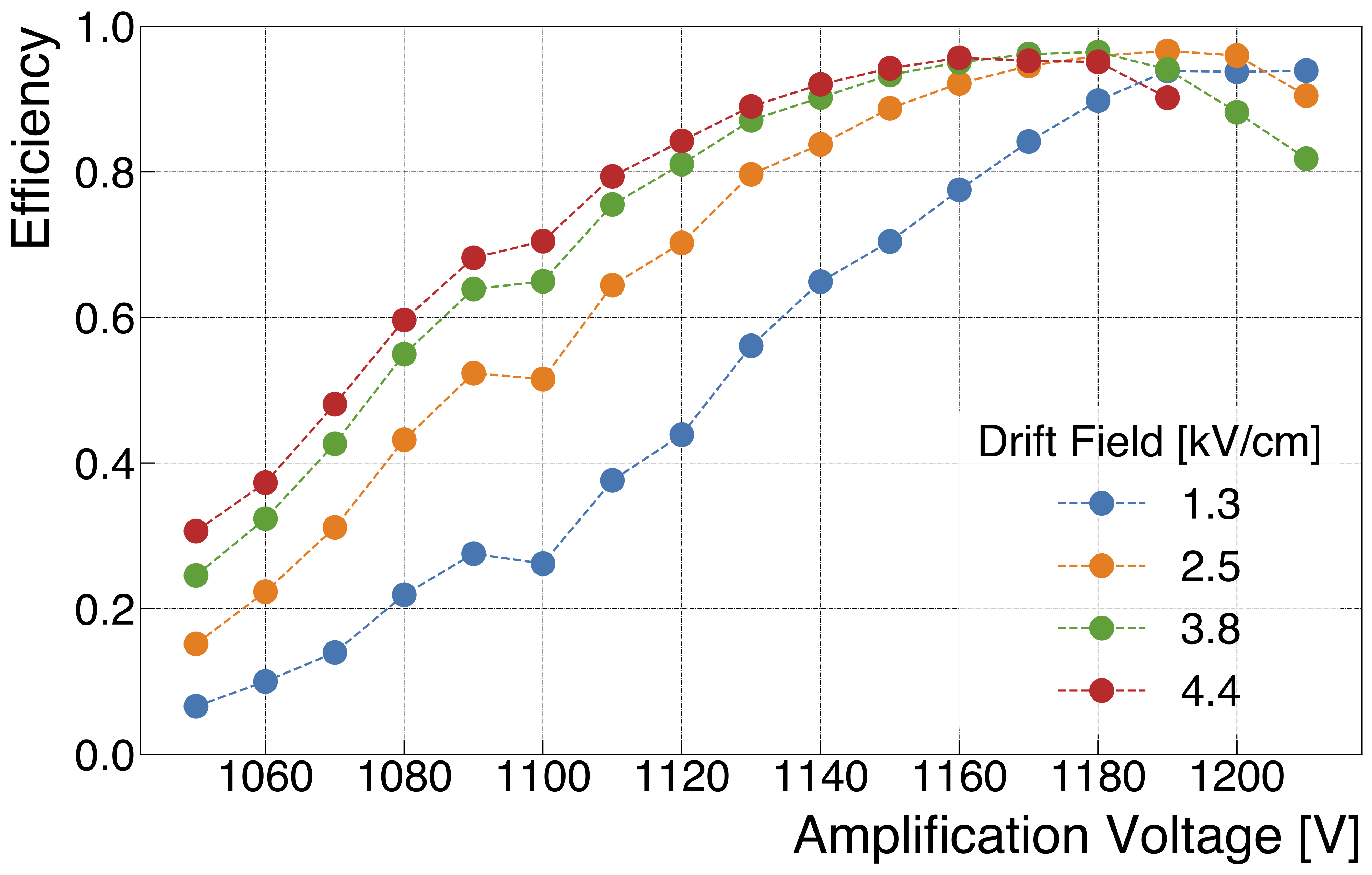}
    \caption{RPWELL}
    \label{fig:efficiency-rpwell}
    \end{subfigure}
    \begin{subfigure}[b]{0.485\textwidth}
    \includegraphics[width=\textwidth]{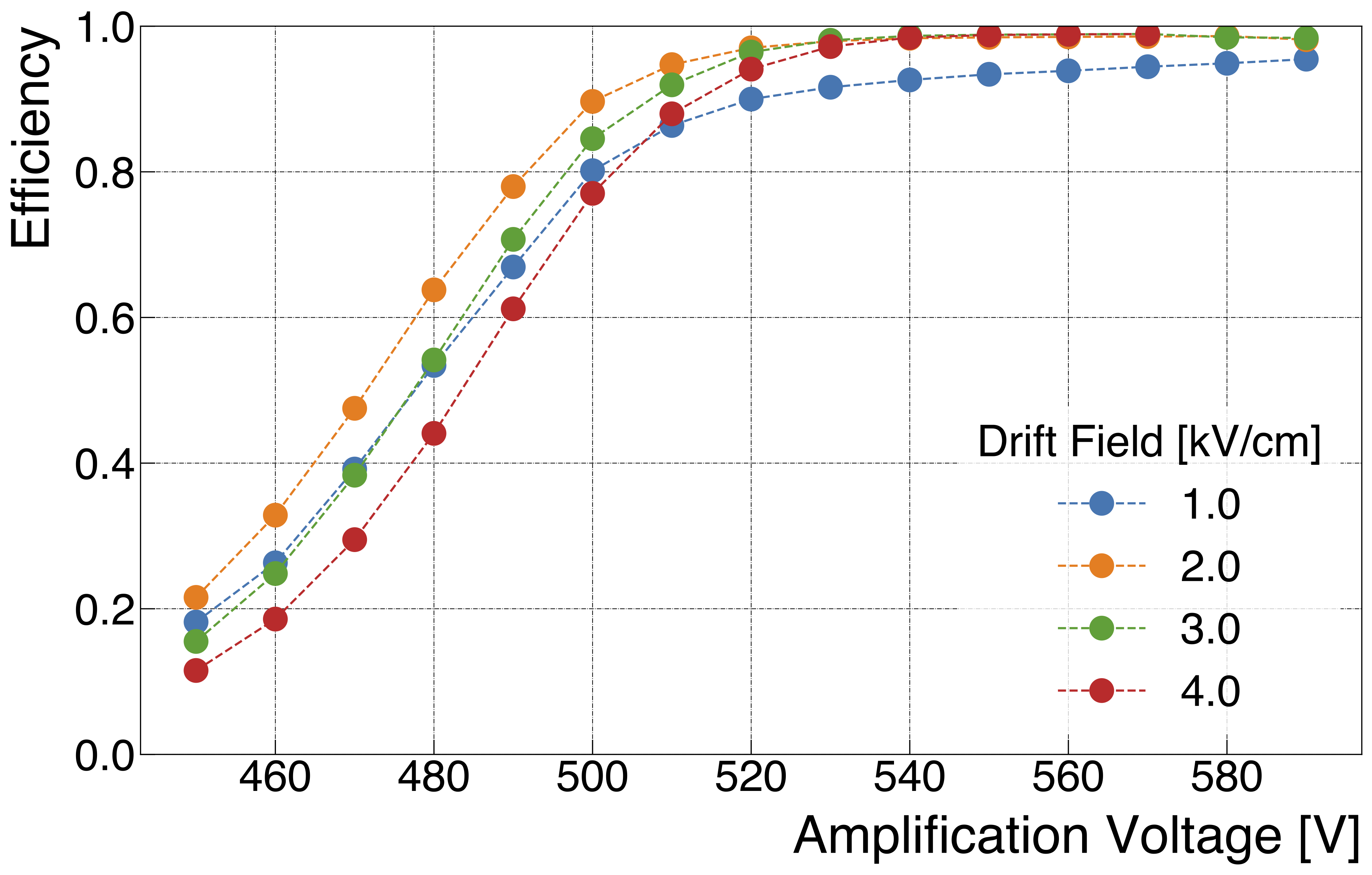}
    \caption{${\mu}$-RWELL}
    \label{fig:efficiency-muRWELL}
    \end{subfigure}
    \caption{\label{fig:efficiency} Efficiency as a function of applied amplification voltage at different drift fields for a) Micromegas, b) RPWELL and c) ${\mu}$-RWELL.}
\end{figure}

An event is selected if it has a reconstructed track. The detector efficiency is defined as a number of events with a cluster matched to the track over the total number of selected events.
The efficiency values as a function of amplification gap voltage for different drift fields are presented in Figure~\ref{fig:efficiency}. For the Micromegas (Figure~\ref{fig:efficiency-mm}), the efficiency of 97\% is achieved for all tested drift field values. However, the higher the drift field value, the higher the amplification gap voltage at which the efficiency plateau is reached. This behavior is attributed to reduced mesh transparency to primary electrons at higher drift fields.

The RPWELL (Figure~\ref{fig:efficiency-rpwell}) operated at the efficiency of 96\% reached at drift field values above 1.3~kV/cm. The moderate slope observed in increasing efficiency is explained by the gradual increase in gain for each step of 10~V. The observed average efficiency drop at extreme gain values is attributed to local regions with higher discharge probability resulting in local efficiency as low as 50\%. 
% VMM3a charge saturation, which occurs with a probability exceeding 10\% at these voltages, compared to only 1–2\% in the rest of the operational range --- for further discussion see \ref{sec:stability}.

An efficiency plateau of 98\% is achieved at drift fields above 2.0~kV/cm with the \urwell (Figure~\ref{fig:efficiency-muRWELL}). As discussed in the previous section, dead areas associated with DLC grounding lines have been excluded from the analysis, leading to significantly improved efficiency compared to the earlier 84\% efficiency plateau reported in \cite{Zavazieva:2025rsj}.

\subsection{Total charge measurement}

The cluster charge distributions, e.g. top row in Figure~\ref{fig:examples}, were fitted to a Landau distribution to extract the Most Probable Value (MPV) as the total charge estimate. The fit is only meaningful when the MPV exceeds the analog and digital thresholds at 440, 1100 and 490~V in the Micromegas, RPWELL and \urwell, respectively. 

\begin{figure}
    \centering
    \begin{subfigure}[b]{0.485\textwidth}
    \includegraphics[width=\textwidth]{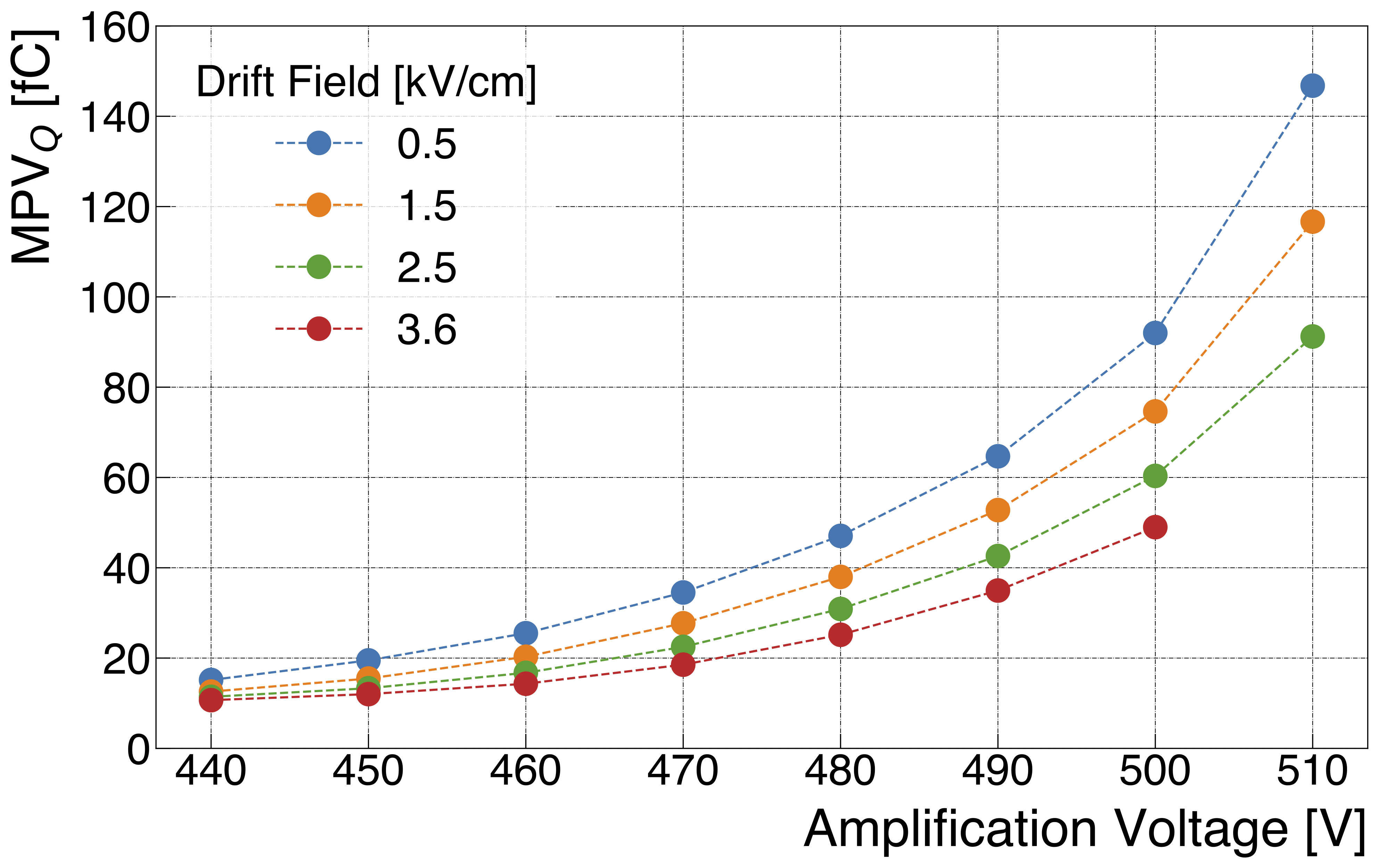}
    \caption{Micromegas}
    \label{fig:q-mm}
    \end{subfigure}
    \begin{subfigure}[b]{0.485\textwidth}
    \includegraphics[width=\textwidth]{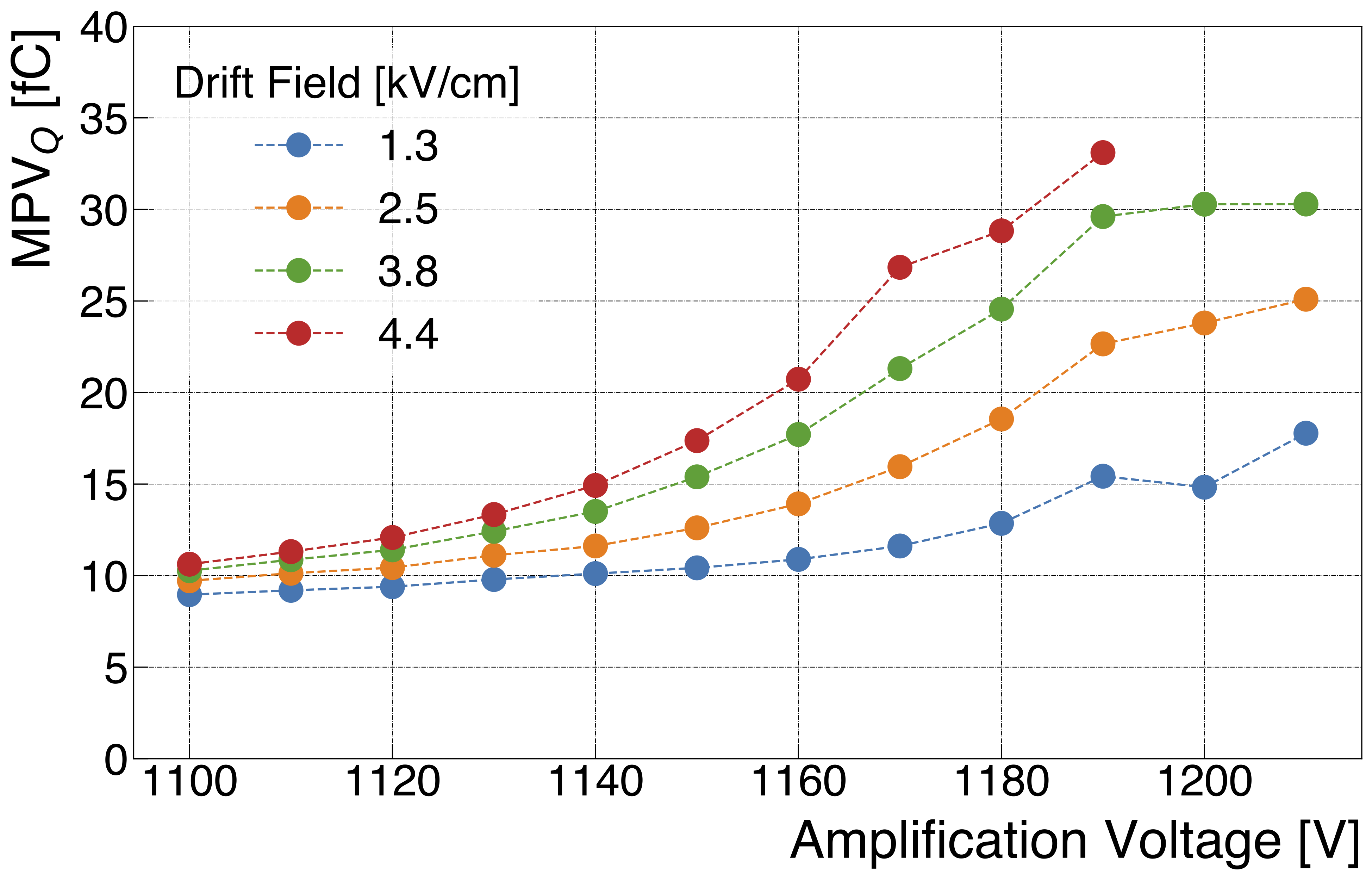}
    \caption{RPWELL}
    \label{fig:q-rpwell}
    \end{subfigure}
    \begin{subfigure}[b]{0.485\textwidth}
    \includegraphics[width=\textwidth]{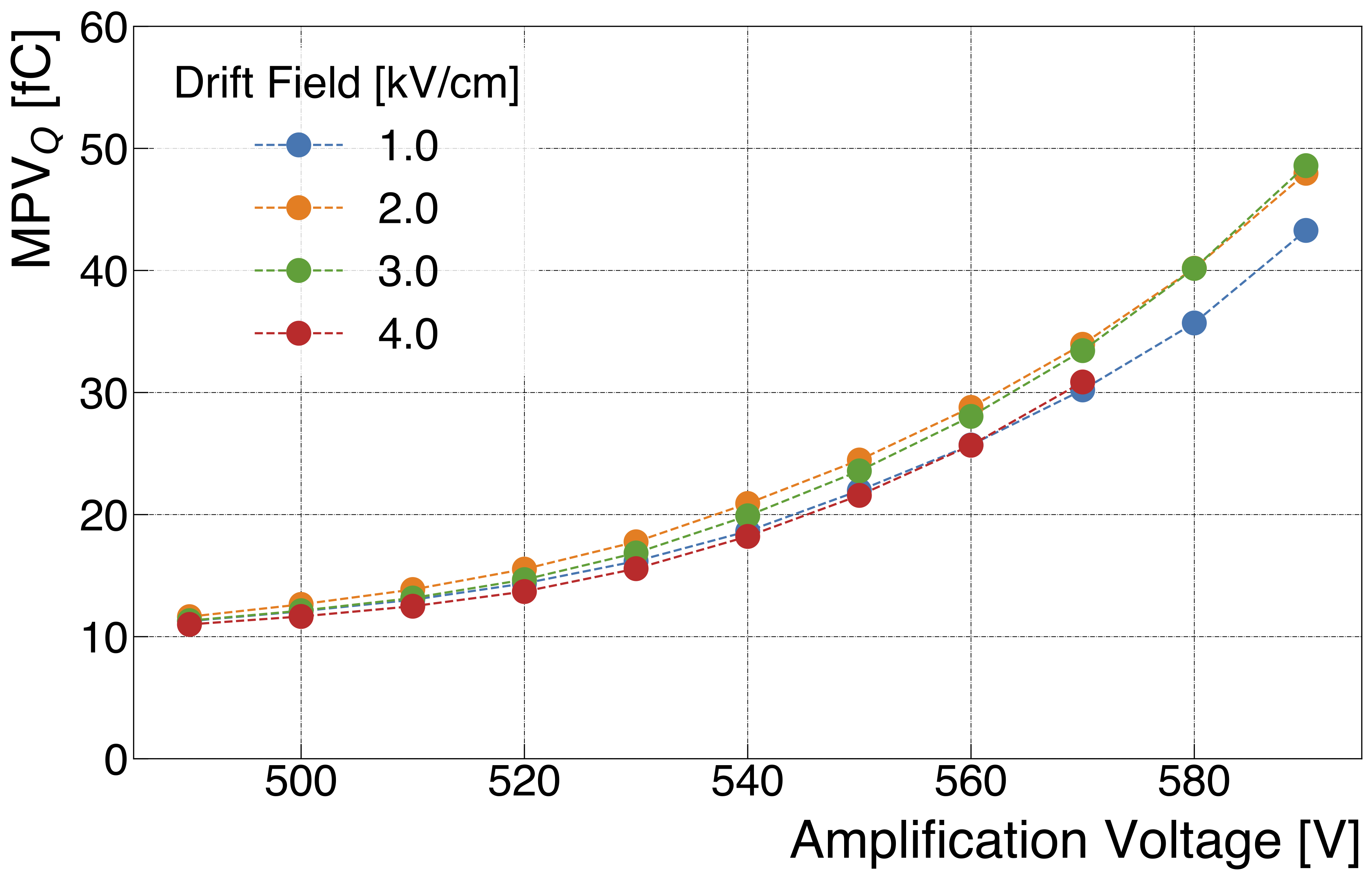}
    \caption{${\mu}$-RWELL}
    \label{fig:q-muRWELL}
    \end{subfigure}
    \caption{The MPV of the charge spectra for the (a) Micromegas, (b) RPWELL, and (c) \urwell. Note the different y-axis scales.}
    \label{fig:q}
\end{figure}

Figure~\ref{fig:q} shows the MPV extracted from the fit as a function of amplification voltage for different drift fields. The Micromegas detector (Figure~\ref{fig:q-mm}) reached the highest charge values indicating high charge gain and full signal integration within the 200~ns peaking time of the VMM3a chip (for details on the signal shape, see Appendix~\ref{app_b}). Full charge integration is also expected in the \urwell (Figure~\ref{fig:q-muRWELL}) due to its short signal rise time. However, the measured charge is lower than that of Micromegas, which is attributed to the overall lower detector gain. % This reduction is primarily due to the lower argon concentration in the gas mixture and the approximately two-times thinner amplification region. 

Due to the long signal rise time of the RPWELL (Figure~\ref{fig:q-rpwell}), only about 30\% of the total charge is integrated with the 200~ns shaping time of the VMM3a. Thus, the absolute gain is roughly three times larger than the reported one. The charge saturation observed at the highest voltages is correlated with the efficiency drop at the same values (see Section \ref{sec:stability}).

\begin{figure}
    \centering
    \begin{subfigure}[b]{0.52\textwidth} 
    \includegraphics[width=\textwidth]{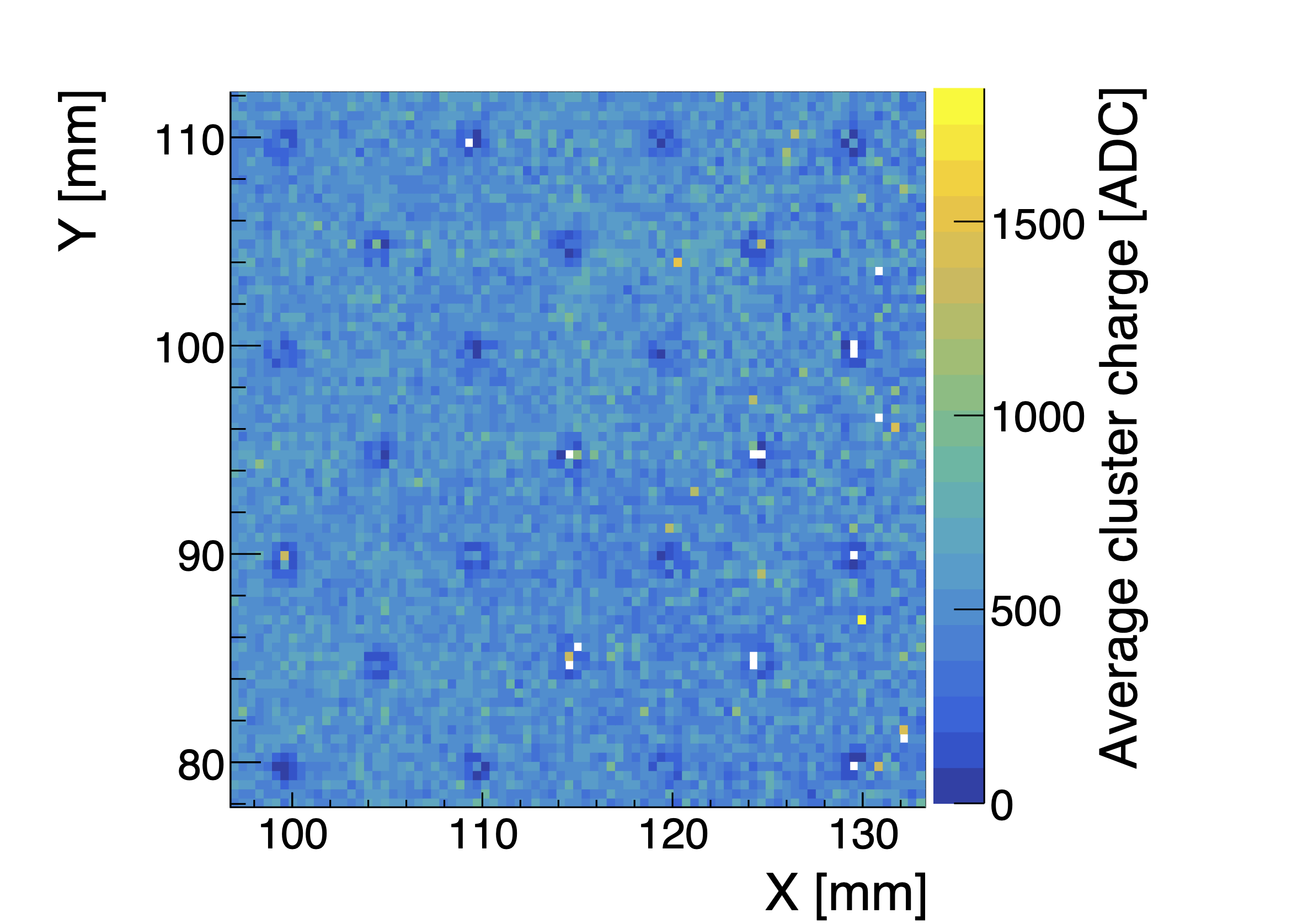}
    \caption{\label{fig:mm_pillars_map}Average cluster charge map}
    \end{subfigure}
    \begin{subfigure}[b]{0.465\textwidth} 
    \includegraphics[width=\textwidth]{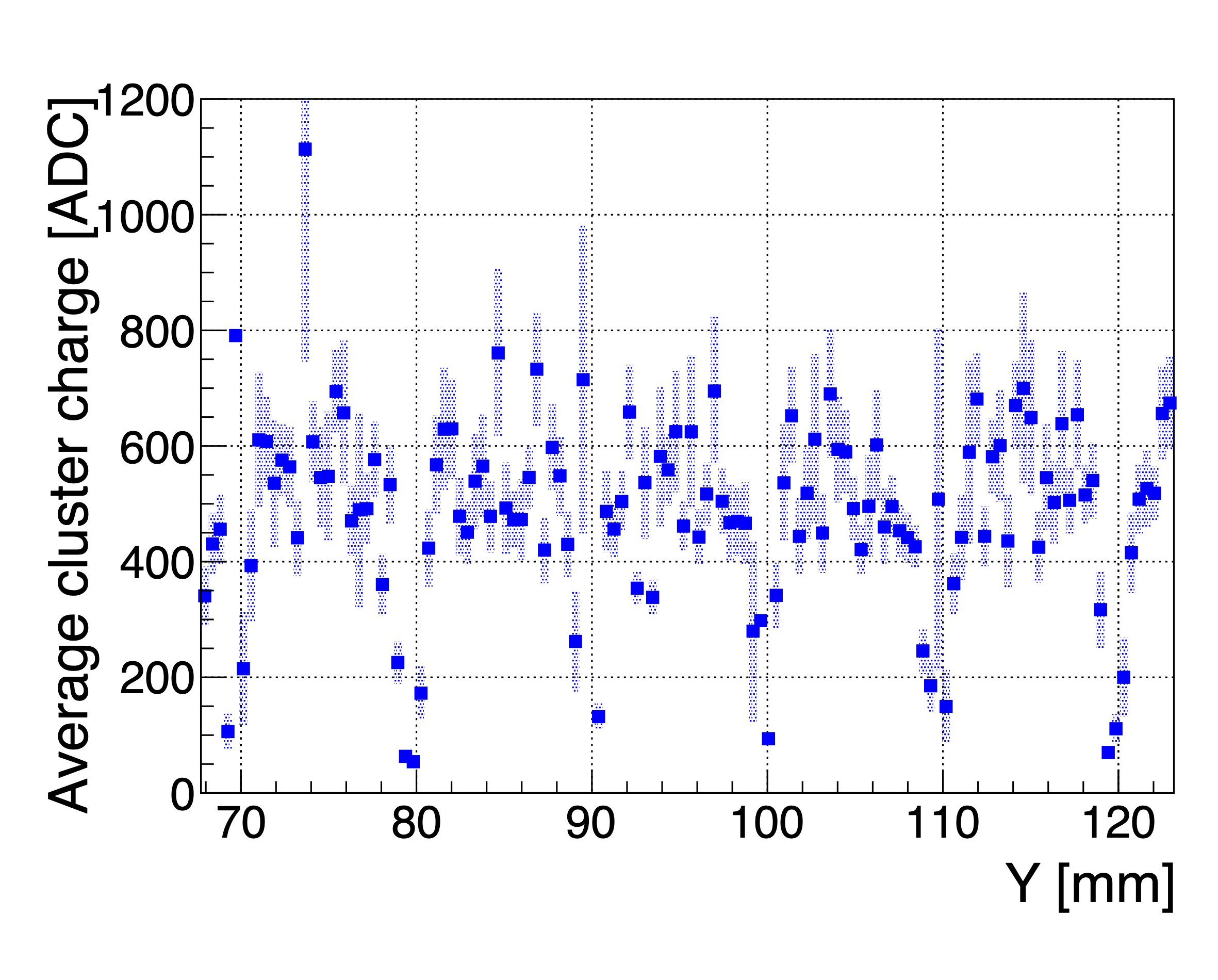}
    \caption{\label{fig:mm_pillars_projection} Map projection along array of pillars}
    \end{subfigure}
    \caption{\label{fig:mm_pillars}Cluster charge measured in Micromegas at 500 V with 0.5 kV/cm drift field: a) map across  $\sim 30 \times 30 ~\mathrm{mm^2}$ of the tested area, b) 1D projection of the map along a column of pillars.}
\end{figure}

The gain uniformity was measured looking at cluster charge maps with a spatial binning of $0.5\times0.5~\mathrm{mm^2}$. The charge map measured with the Micromegas is shown in Figure \ref{fig:mm_pillars_map}. A zoom into $30\times30~\mathrm{mm^2}$ segment is shown for improved visibility. The average charge in the vicinity of the support pillars drops to about half of that in the regions between the pillars --- Figure \ref{fig:mm_pillars_projection} --- slicing a $0.5 \times 60~\mathrm{mm^2}$ strip ($95.5~\mathrm{mm}<\mathrm{X}<100 ~\mathrm{mm}$ and $70~\mathrm{mm}<\mathrm{Y}<130 ~\mathrm{mm}$) across seven pillars.

\begin{figure}
    \centering
    \begin{subfigure}[b]{0.43\textwidth}
    \includegraphics[width=\textwidth]{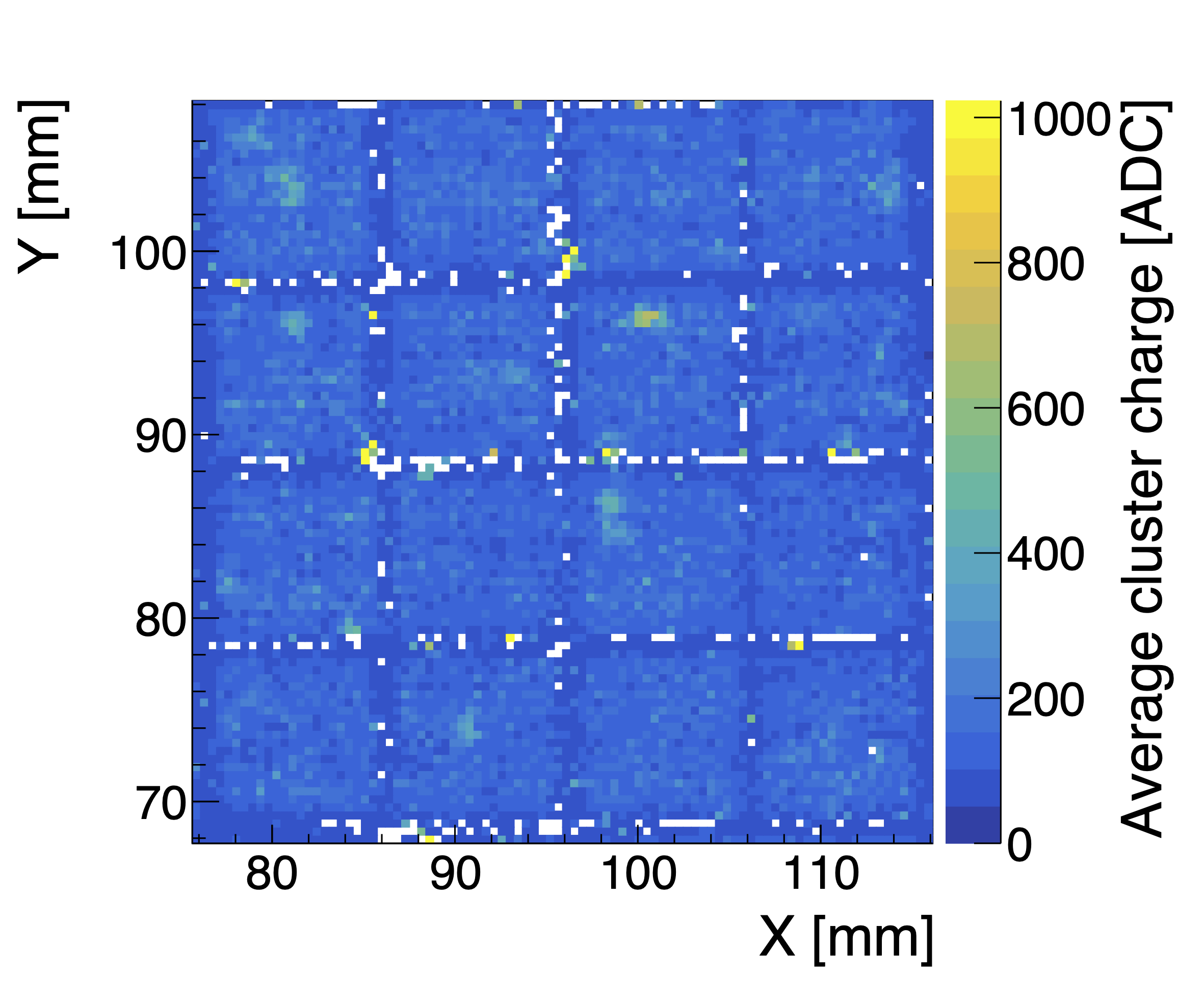}
    \caption{Charge of size = 1 clusters}
    \label{fig:rpwell_m1_map}
    \end{subfigure}
    \begin{subfigure}[b]{0.43\textwidth}
    \includegraphics[width=\textwidth]{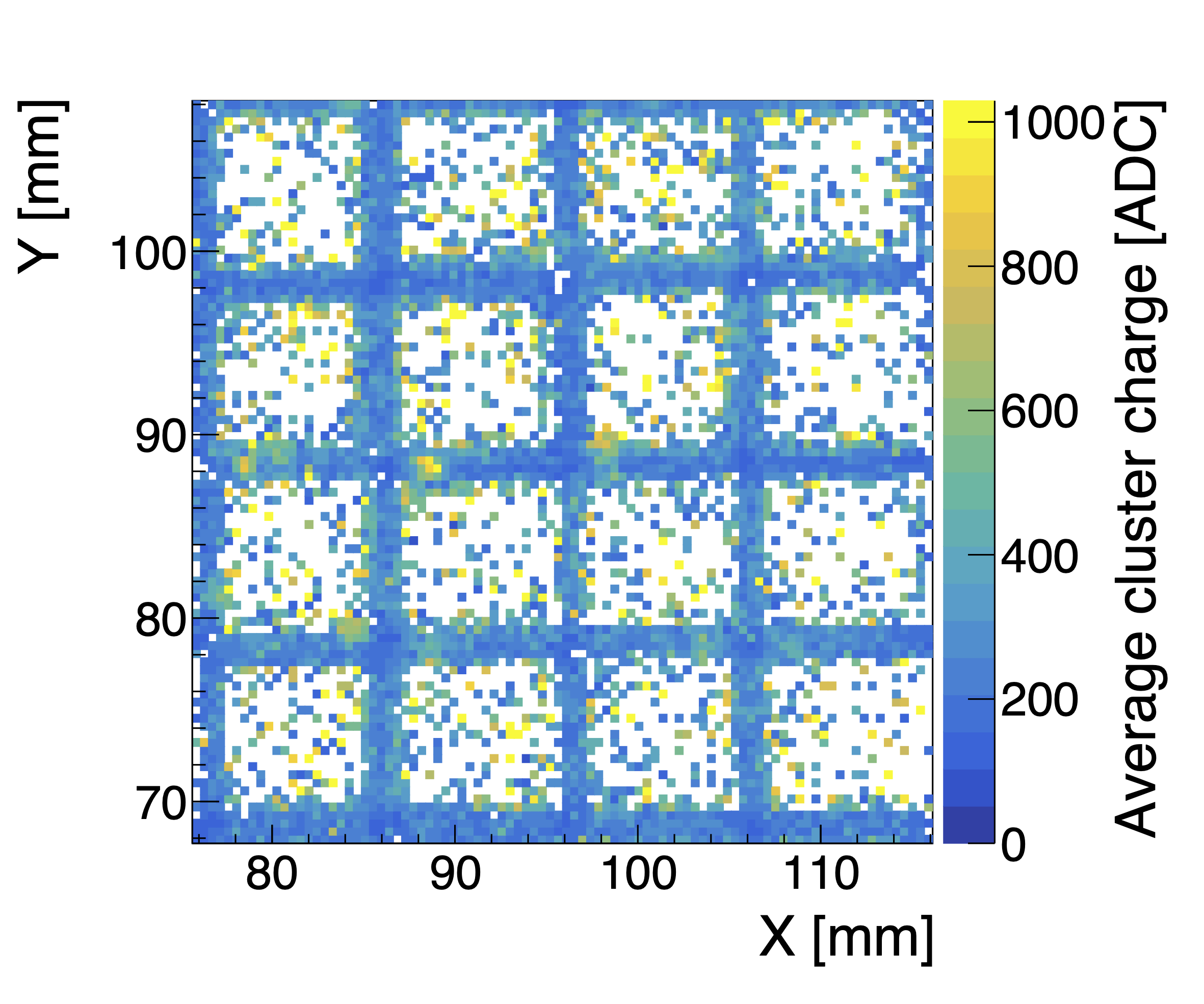}
    \caption{Charge of size = 2 clusters}
    \label{fig:rpwell_m2_map}
    \end{subfigure}
    \\
    \begin{subfigure}[b]{0.56\textwidth}
    \includegraphics[width=\textwidth]{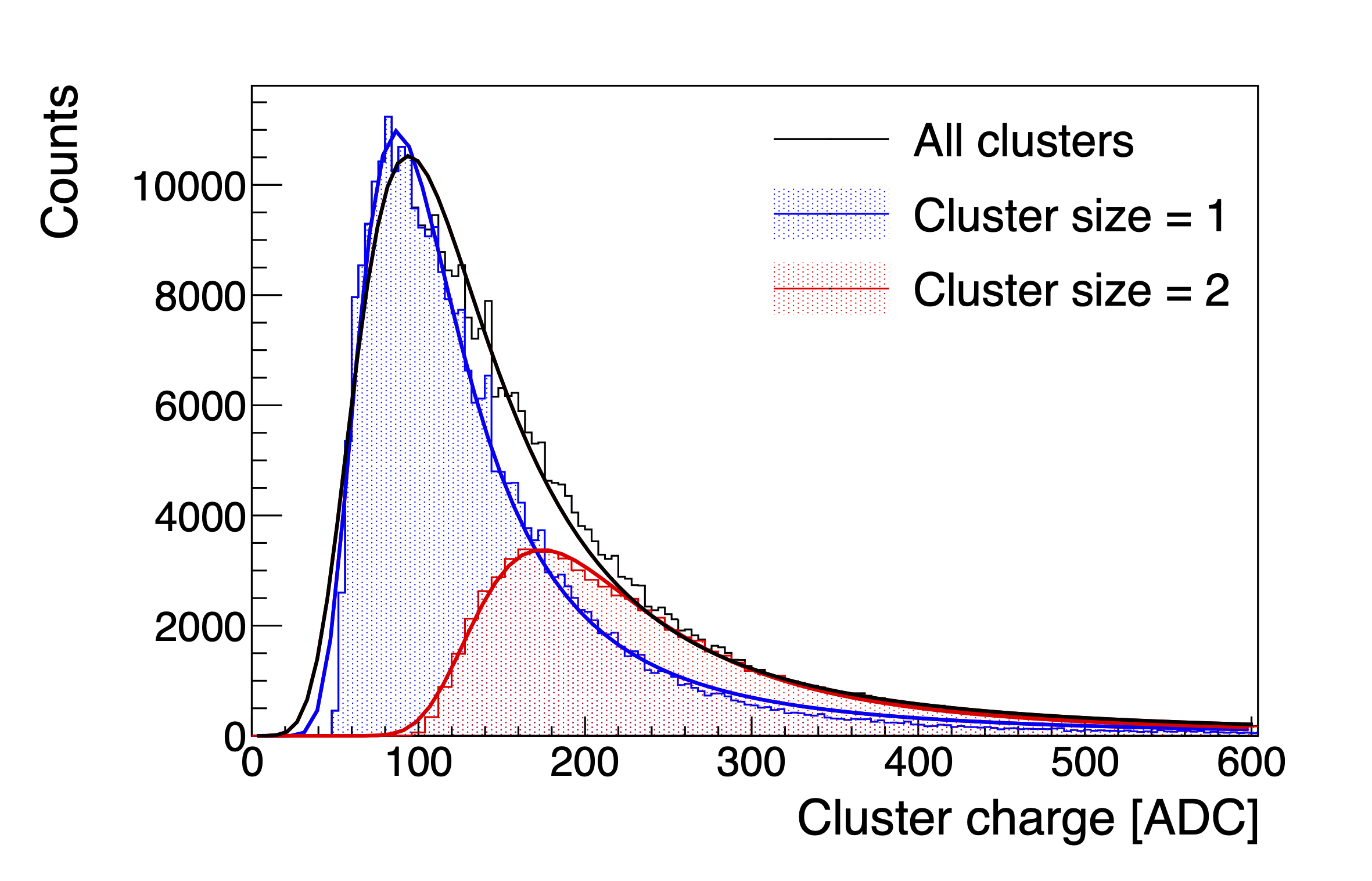}
    \caption{Cluster charge distributions}
    \label{fig:rpwell_m1_m2_all}
    \end{subfigure}
    \caption{Cluster charge measured in RPWELL at 1190V with 2.5 kV/cm drift field: a) charge map of size = 1 clusters, b) charge map of size = 2 clusters, c) comparison of charge distributions distinguishing the two cluster sizes, solid lines represent the fit to Landau function.}
    \label{fig:rpwell-edge}
\end{figure}

Figure~\ref{fig:rpwell-edge} shows segments of  $40\times40 ~\mathrm{mm^2}$ maps of the average RPWELL cluster charge; for cluster size 1 (Figure \ref{fig:rpwell_m1_map}) and 2 (Figure \ref{fig:rpwell_m2_map}). Figure \ref{fig:rpwell_m1_m2_all} shows three cluster charge distributions: clusters of size 1, clusters of size 2, and the overall cluster charge. A clear grid pattern of $10\times10 ~\mathrm{mm^2}$ across the segment is seen. This grid corresponds to the segmentation of both the readout board and the THGEM top electrode. While the larger cluster multiplicity (see Figure \ref{fig:examples}) is attributed to the former, the higher gain is attributed to the stronger amplification field close to the conductive strips defining the top electrode segmentation. Similar phenomena were studied thoroughly in the context of COMPASS RICH detector~\cite{ALEXEEV201796} and was not observed in the Micromegas and the \urwell.

\begin{figure}
    \centering
    \includegraphics[width=0.5\linewidth]{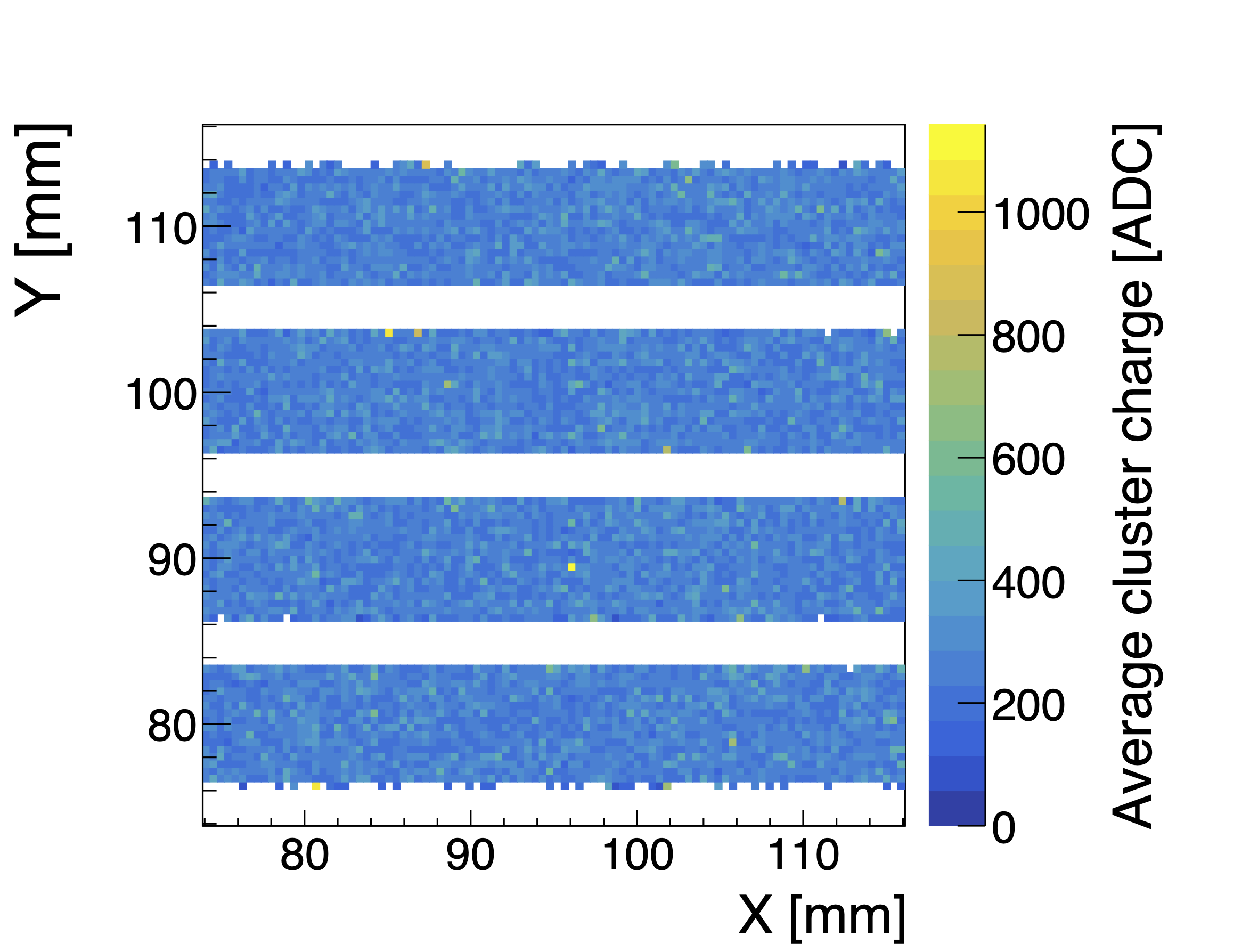}
    \caption{Cluster charge measured in \urwell at 590~V with 2.0 kV/cm drift field.}
    \label{fig:urwell_map_no_dead_area}
\end{figure}

The \urwell detector demonstrated overall good performance in terms of gain uniformity across the evaluated area, excluding approximately 30\% of the surface affected by the DLC grounding lines. The average cluster charge map for the \urwell (excluding dead regions) is shown in Figure~\ref{fig:urwell_map_no_dead_area}.

\begin{figure}
    \centering
    \includegraphics[width=0.55\linewidth]{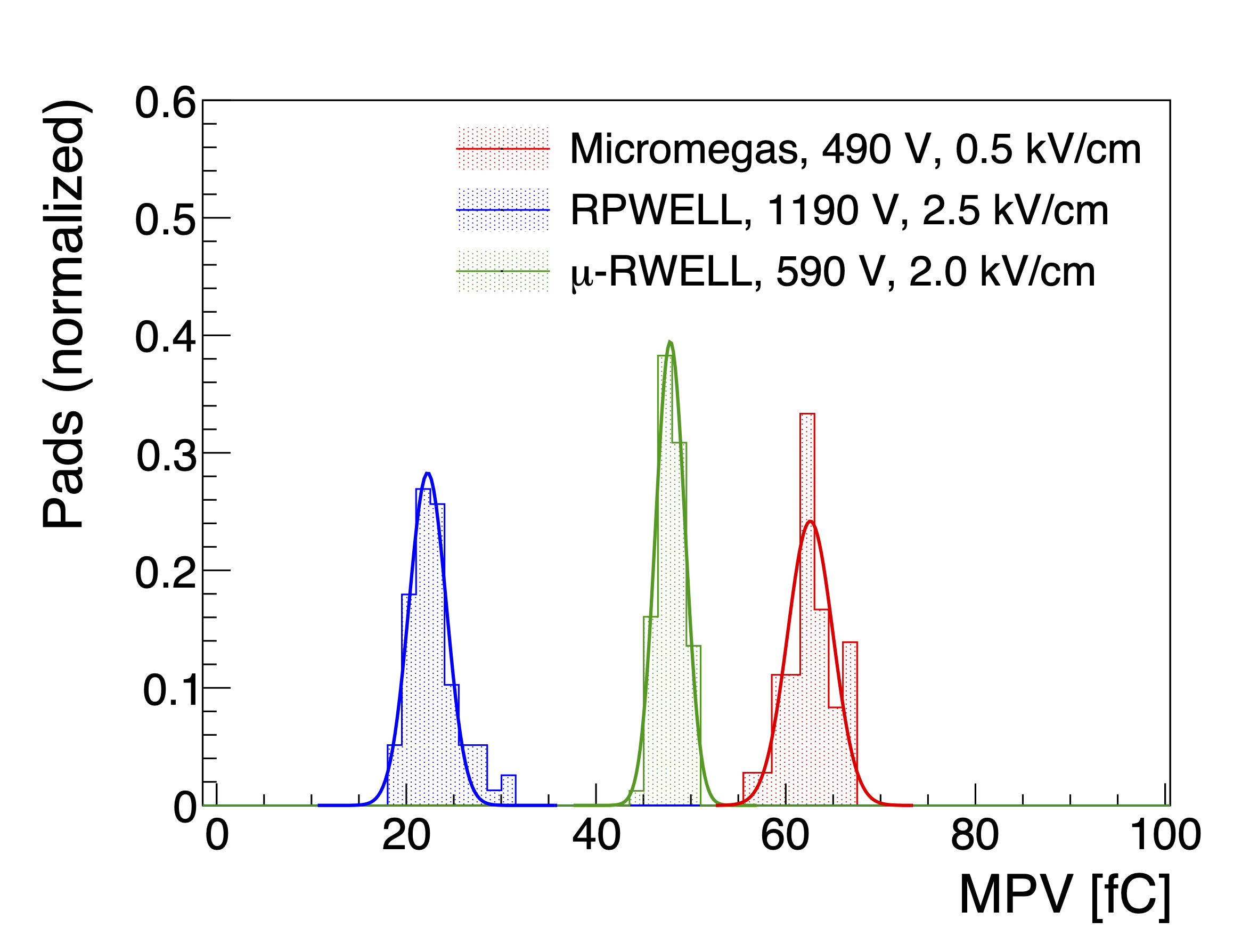}
    \caption{Cluster charge MPV (from fit to Landau distribution) across 64 readout pads.}
    \label{fig:q_uniformity}
\end{figure}

Figure~\ref{fig:q_uniformity} shows a comparison of MPV distributions of the three technologies. The data was collected at the efficiency plateau prior to the occurrence of instabilities or saturation effect at 490, 1190 and 590~V in the Micromegas, RPWELL and \urwell, respectively. A sufficient statistics enabled the analysis across $8\times8$ pads. The gain uniformity is quantified as the ratio $\sigma/\mu$ from a Gaussian fit. A uniformity of 2.8\%, was measured with the \urwell prototype, consistent with a qualitative observation of average cluster charge maps above. A slightly poorer value of 3.4\% was obtained for the Micromegas. It should be noted that this estimate does not account for local non-uniformities within a single pad area, introduced by the pillars. Since each pad contains one pillar at its center and approximately 25\% of a pillar at each corner, the effect is similar for all pads and does not contribute to pad-to-pad differences.
For the RPWELL detector, a uniformity of 6.7\% was recorded\footnote{ This value might improve if the full charge integration is performed}. This higher value reflects gain variations stemming from thickness non-uniformity of the multiplier electrode, as well as non uniform attachment of the electrode to the resistive plate — secured only at discrete points separated by distances of at least 5 cm~\cite{zavazieva2023towards}.

\subsection{Electrical stability}
\label{sec:stability}

\begin{figure}
    \centering
    \includegraphics[width=0.65\textwidth]{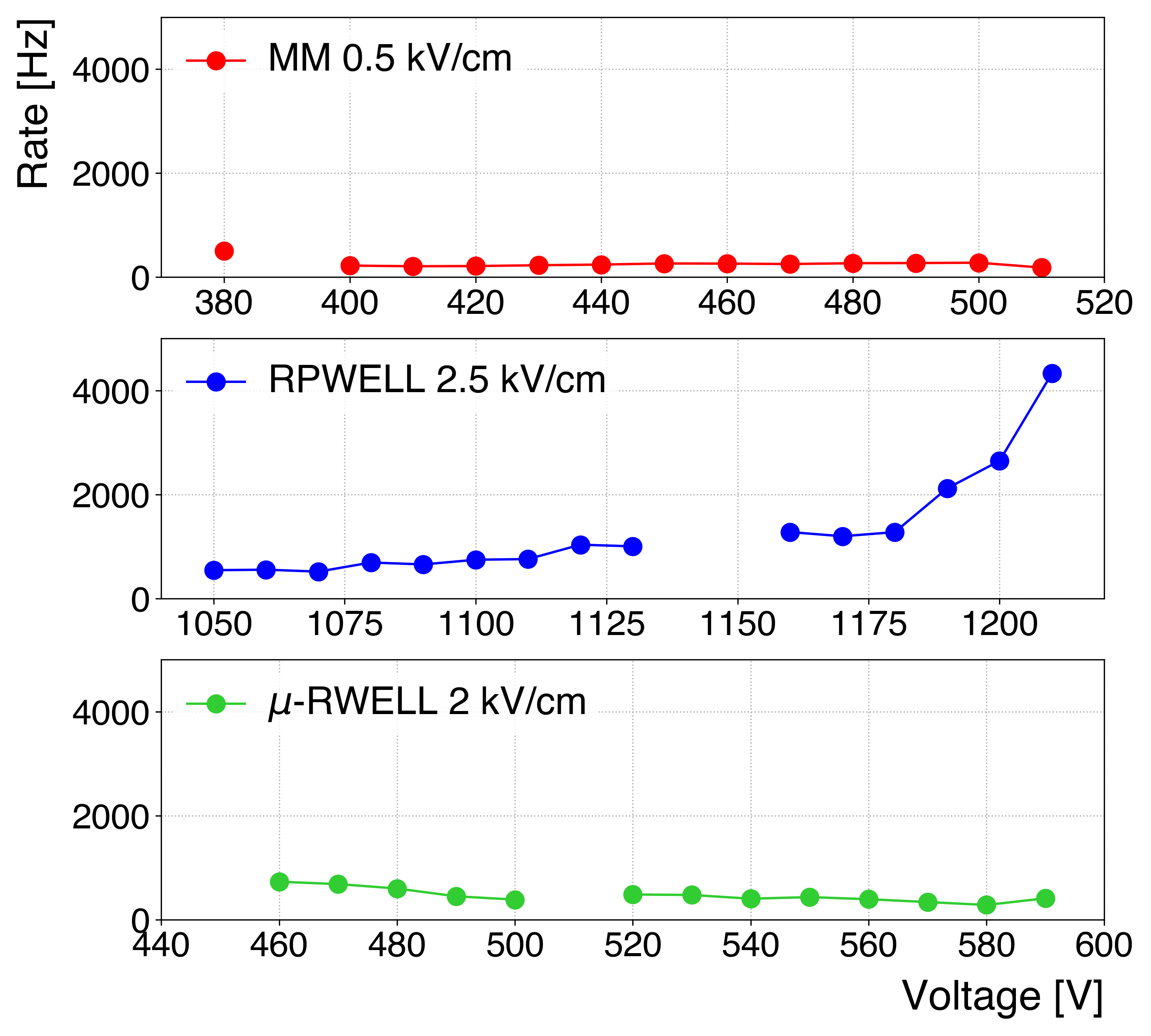}
    \caption{\label{fig:cluster_rates_off_spill} Cluster rate recorded with the tested detectors in between beam spills. Missing data points correspond to runs affected by external noise.}
\end{figure}

Typically, a single run consists of 6 to 9 spills of 4.5 seconds separated by 10 seconds intervals with no traversing particles\footnote{ A low rate of individual particles still present, not triggering the system}. Figure \ref{fig:cluster_rates_off_spill} shows the rate of clusters recorded in the 10-seconds intervals in between spills as a function of applied amplification voltage. Their typical charge values are low, but higher values are recorded as well including occasional saturated signals. Their origin can be attributed to pick-up noise, cosmic muons, beam halo and intrinsic detectors spurious activity. Notably, in the RPWELL detector, the rate increases with voltage, reaching $4\cdot10^3$ Hz at the highest applied voltage. This behavior suggests detector instability component dominating over the noise.

\begin{figure}
    \centering
    \includegraphics[width=0.65\textwidth]{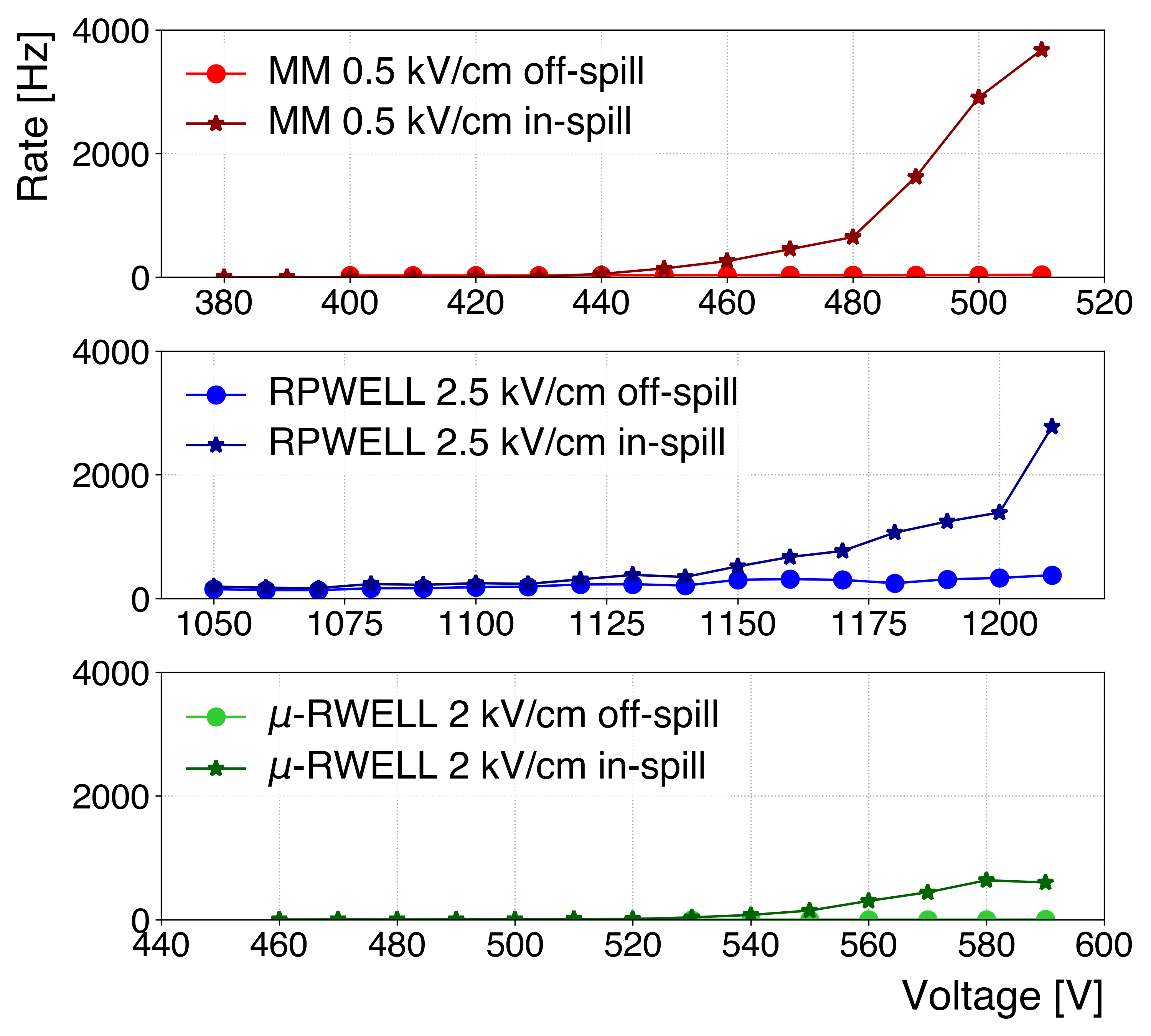}
    \caption{\label{fig:discharge_rate} Discharge rate as a function of applied voltage for the three detectors.}
\end{figure}

Figure~\ref{fig:discharge_rate} shows the rate of discharge-like events measured both during beam spills and in the off-beam intervals. Baseline off-beam discharge rate of approximately 10, $10^2$ and 1 Hz is measured in the Micromegas, RPWELL and \urwell, respectively across all applied voltages.
In-beam, all three detectors show an increase in discharge rate with rising gain. For Micromegas, the rate increases from 450 Hz to 3700 Hz across the efficiency plateau (for voltages above 470 V), corresponding to roughly a threefold gain increase. In the RPWELL, discharge activity starts at 1000 Hz at the onset of the efficiency plateau (1180 V) and rises to 2800 Hz at the highest applied voltage, where an efficiency drop is observed. The $\mu$-RWELL shows the lowest overall discharge activity, with rates increasing from 11 to 600 Hz over the 520–590 V efficiency plateau.

% As for the intensity of discharges, the RPWELL detector shows much larger clusters with saturated pad(s).

\subsection{Time resolution}

\begin{figure}
    \centering
    \includegraphics[width=0.6\textwidth]{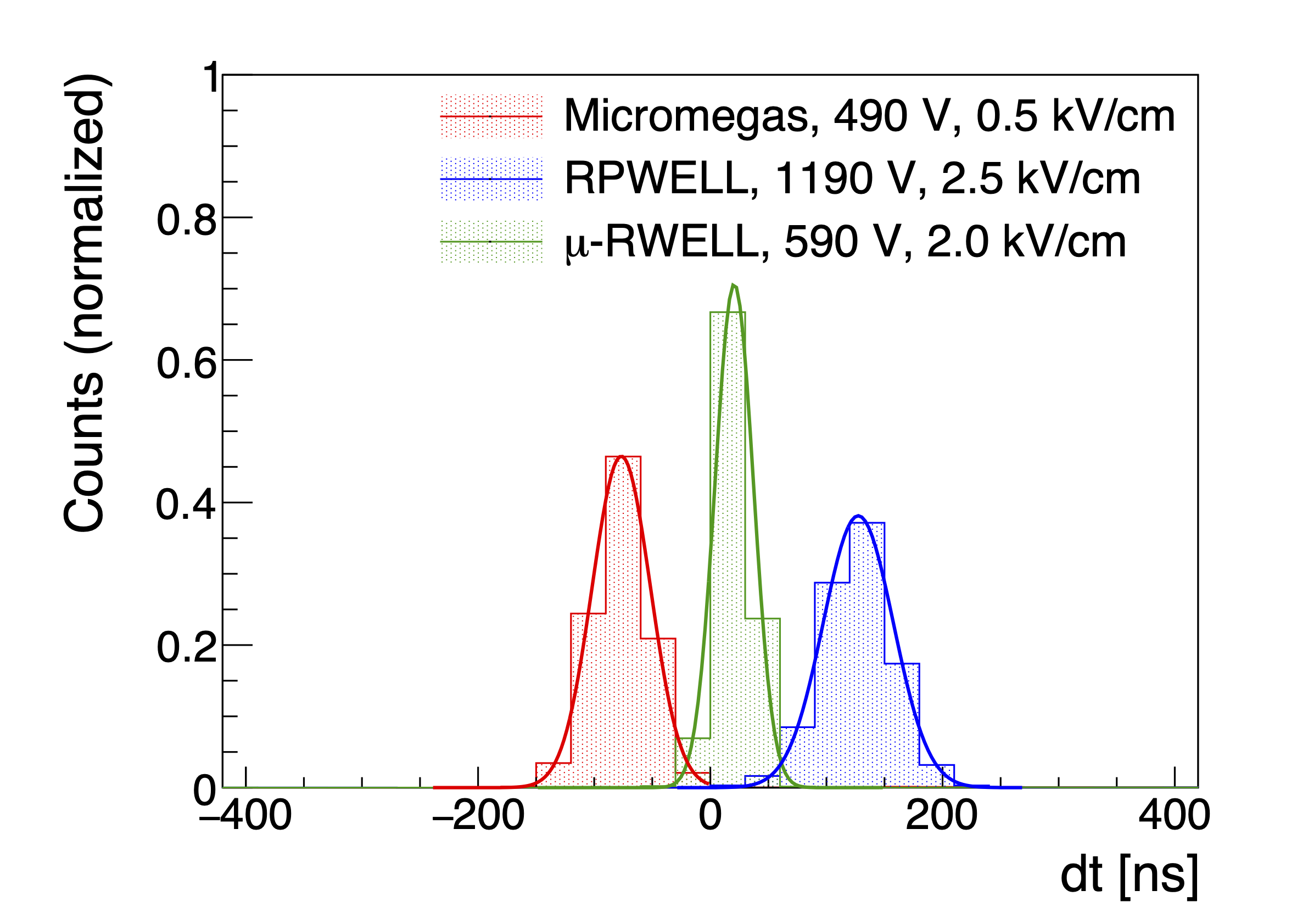}
    \caption{\label{fig:dt_examples} Typical dt distributions measured with the tested detectors.}
\end{figure}

\begin{figure}
    \centering
    \begin{subfigure}[b]{0.485\textwidth}
    \includegraphics[width=\textwidth]{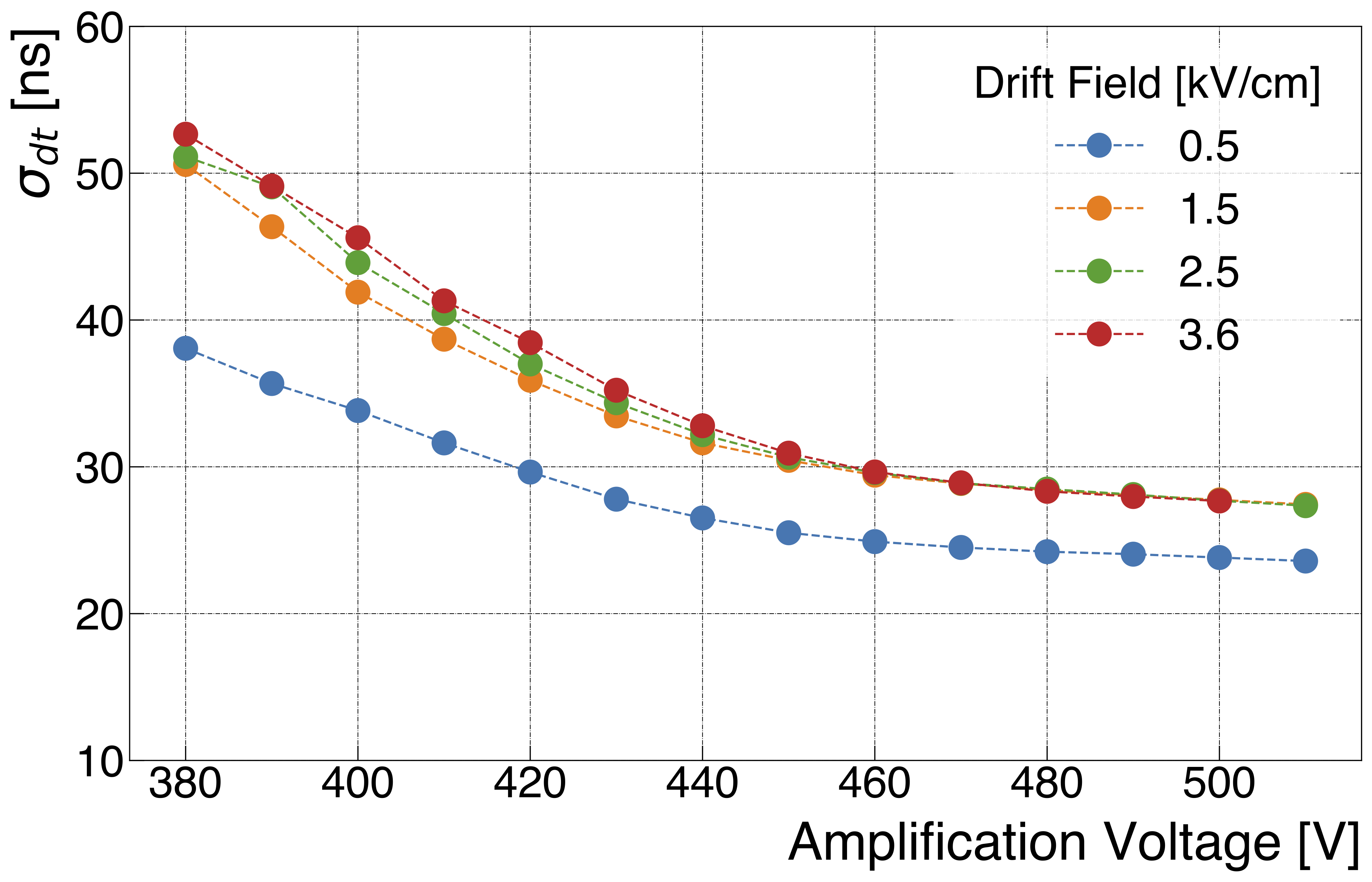}
    \caption{Micromegas}
    \label{fig:t-mm}
    \end{subfigure}
    \begin{subfigure}[b]{0.485\textwidth}
    \includegraphics[width=\textwidth]{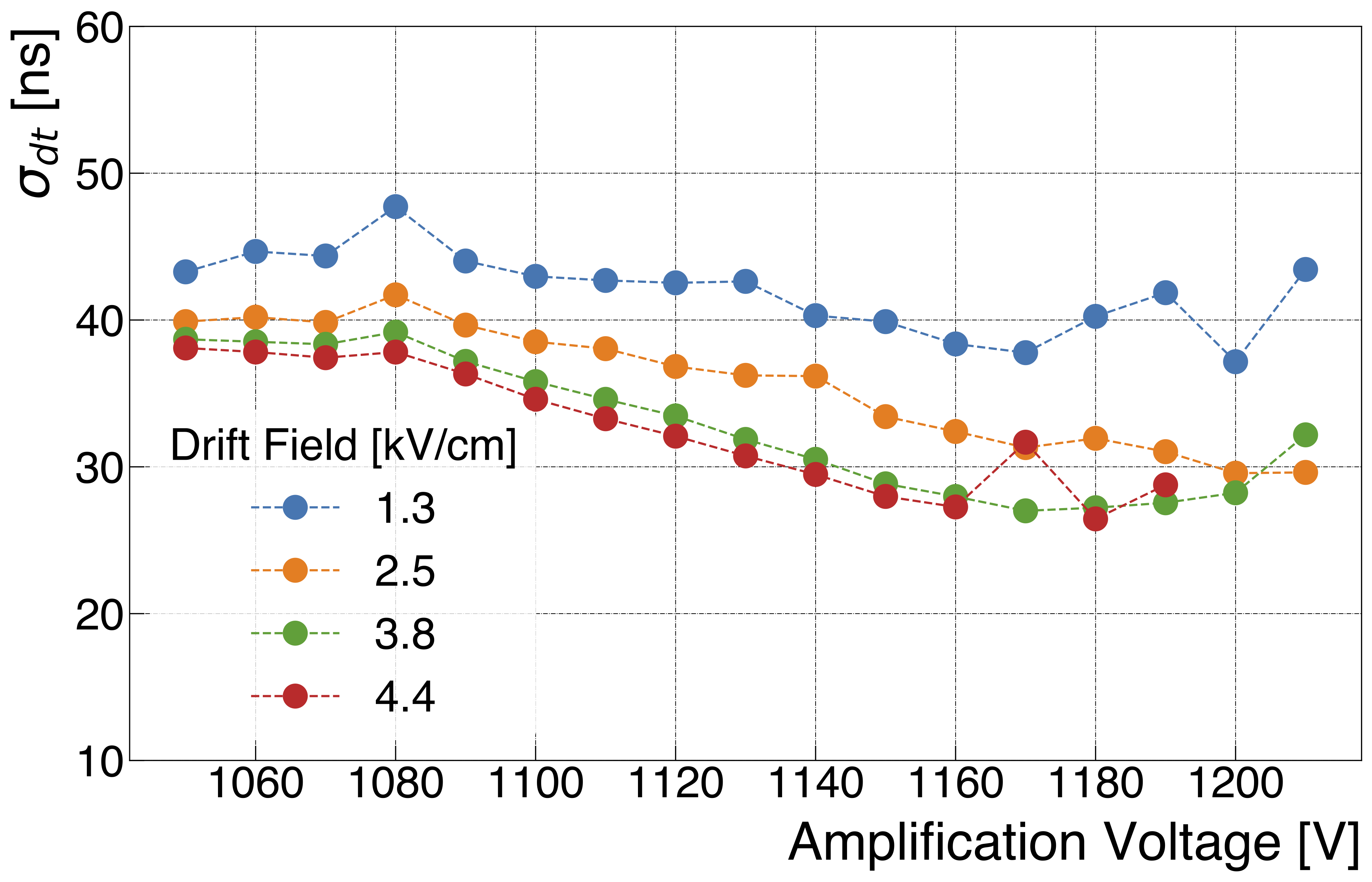}
    \caption{RPWELL}
    \label{fig:t-rpwell}
    \end{subfigure}
    \begin{subfigure}[b]{0.485\textwidth}
    \includegraphics[width=\textwidth]{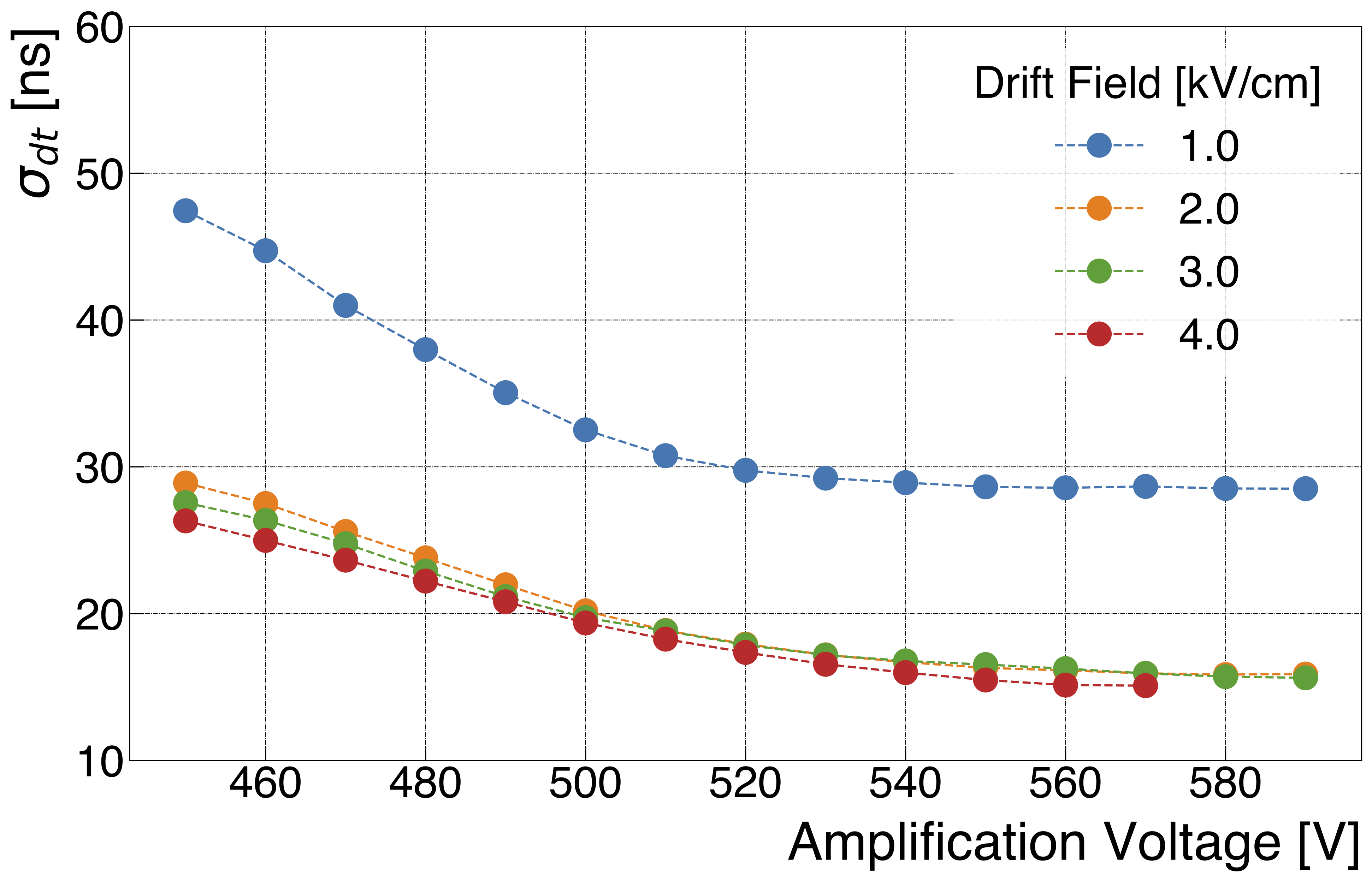}
    \caption{${\mu}$-RWELL}
    \label{fig:t-muRWELL}
    \end{subfigure}
    \caption{\label{fig:time} Time resolution of the DUT with respect to the scintillator trigger signal: a) Micromegas, b) RPWELL  and c) ${\mu}$-RWELL.}
\end{figure}

Figure~\ref{fig:dt_examples} shows typical $\mathrm{dt}$ distributions of the Micromegas, RPWELL and \urwell operated at the efficiency plateau prior to the occurrence of instabilities, at 490, 1190 and 590 V, respectively. Each distribution was fitted with a Gaussian function. The width of the fit ($\sigma$) is defined as the time resolution and is summarized in Figure~\ref{fig:time}. The measured peak time values incorporate contributions from various components of the setup: intrinsic time jitter of the gaseous detectors, the VMM3a ASIC clock and time-to-digital converter (TDC) resolution, firmware offsets in the SRS Front-End Concentrator (FEC), and delays introduced by the NIM discriminator and coincidence logic unit.

The RPWELL, featuring a thicker amplification gap and operated with a slower gas mixture, exhibited larger time fluctuations, resulting in degraded time resolution. Micromegas, which used the same gas mixture as RPWELL but with an amplification gap four times thinner, showed a slightly better resolution. Both Micromegas and RPWELL demonstrated similar time resolution at their respective optimal drift fields, 25 ns relative to 15 ns measured with the $\mu$-RWELL. The superior timing of the $\mu$-RWELL is attributed to its thinner amplification gap and the use of a faster gas mixture. Notably, the time resolution of Micromegas deteriorated with increasing drift field, while in both the $\mu$-RWELL and RPWELL, it improved.

\subsection{Rate capability}

\begin{figure}
    \centering
    \includegraphics[width=0.6\linewidth]{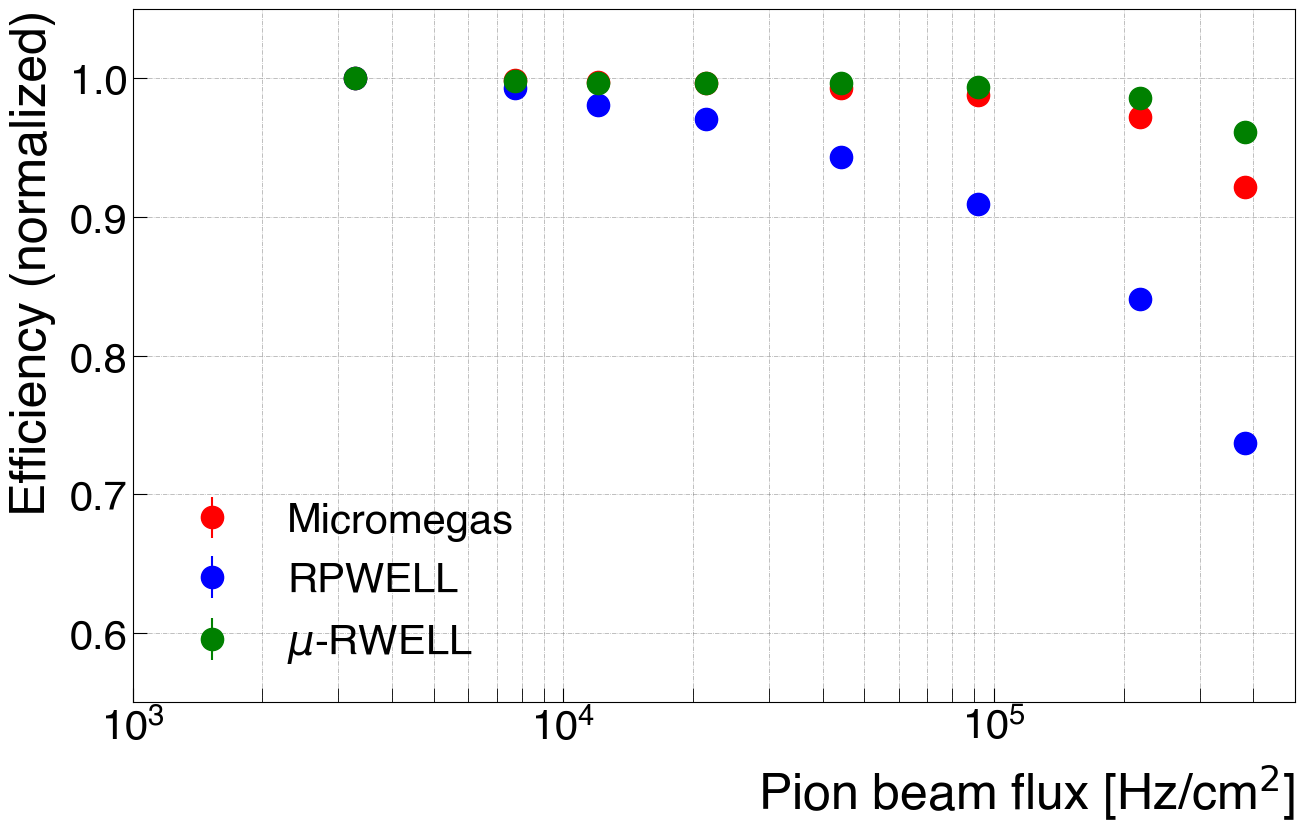}
    \caption{Efficiency of the tested detectors as a function of pion beam flux. }
    \label{fig:pion-rate-scan}
\end{figure}

The efficiency as a function of incoming pion beam flux is shown in Figure~\ref{fig:pion-rate-scan}. The detectors were operated at their efficiency plateau, and the efficiency values were normalized to 1 at the lowest flux point of 3~kHz/cm$^2$.

Having the highest resistivity (2 ~$\mathrm{G}\Omega$), the RPWELL detector exhibits a more pronounced performance degradation at higher rates. Its efficiency starts to drop significantly around $10^4$~Hz/cm$^2$, decreasing to approximately 74\% at $10^5$~Hz/cm$^2$. 
The resistive layer with a significantly lower surface resistivity of 50–100~M$\Omega/\Box$ in the \urwell  and the Micromegas enables the detectors to maintain higher efficiencies of 92\% and 96\%, respectively, over the same flux range.
    
\section{Discussion and outlook}
\label{sec:discussion}

This paper presents a comparison of three resistive MPGD technologies --- Micromegas, RPWELL and \urwell --- using relativistic muons and pions. Based on past experience, the Micromegas and RPWELL were operated with $\mathrm{Ar/CO_2/iC_4H_{10}}$, while the \urwell with greenhouse $\mathrm{Ar/CO_2/CF_4}$ gas mixture.

After excluding from the analysis detection areas with known design problems and dead electronic channels, high detection efficiency of 97\%, 96\% and 98\% was measured with the Micromegas, RPWELL and \urwell, respectively.

Higher charge values, up to 143 fC, were measured with the Micromegas relative to 35 fC and 50 ADC measured with the RPWELL and \urwell, respectively. For the RPWELL, it represents only about 30\% of the total avalanche charge due to a mismatch between the VMM3a 200 ns integration time and 1 $\mu$s signal rise time.  Average pad charge uniformity of 3.4\% was measured in the Micromegas. Within each pad, the charge measured up to 2 mm around the pillars was about half of that measured farther away. In the RPWELL, the average pad uniformity was 6.7\%. Higher charge values were consistently measured close to the pad boundaries of the THGEM holes' pattern. The average pad uniformity in the \urwell was 2.8\%.

Clusters were recorded in all three detectors also in the absence of the beam. A constant rate of about  $10^2$ Hz (across the entire active area) was measured both in the Micromegas and \urwell detectors. In the RPWELL, the rate increased with the voltage, reaching a maximum of $10^3$ Hz. Discharge rate was recorded off- and in-beam time intervals. Off-beam, in the Micromegas, RPWELL and \urwell 10 Hz, $10^2$ Hz and 1 Hz were measured, respectively. In-beam, the discharge rate in the three detectors increased with operation voltage, reaching $10^3$ Hz in the Micromegas and RPWELL, and $10^2$ Hz in the \urwell. 
 
Approximately 25 ns time resolution could be reached both with the Micromegas and RPWELL, while values at the order of 15 ns were measured in the \urwell. The rate capability of Micromegas and \urwell is similar, with almost no efficiency loss up to incoming particle rates of 100~kHz/cm$^2$, whereas the efficiency of the RPWELL slowly drops from about 10~kHz/cm$^2$, consistent with the difference in resistivity between the tested detectors.

Equipped with 1 $\mathrm{cm^2}$ pad readout, the three technologies are promising for moderate precision tracking, providing high detection efficiency and resilience to discharges. The \urwell and Micromegas are more suitable for high-rate environments.

Future work should attempt to optimize the resistivity of the RPWELL, resolve the dead areas problem and search for environment friendly gas mixture for the \urwell. Off-beam activities, highest in the RPWELL, should be minimized. Scaling-up these detectors is also foreseen.

\acknowledgments

This study is supported by the Minerva Foundation with funding from the Federal German Ministry for Education and Research, as well as by the Krenter-Perinot Center for High-Energy Particle Physics. Additional support comes from a research grant provided by Shimon and Golde Picker, the Nella and Leon Benoziyo Center for High Energy Physics, and the Sir Charles Clore Prize. Special thanks go to Martin Kushner Schnur for his invaluable support to this research.

\appendix

\section{Signal Shape}
\label{app_b}

In Micromegas, signals are primarily generated by the drift of ions across the amplification gap, resulting in a rise time of $\sim$150–200~ns. A very fast electron component, representing about 10–15\% of the total induced charge, can in principle be detected using a fast current-sensitive preamplifier with a rise time below 100~ps~\cite{Bortfeldt:2010development}. 

The electron component in the \urwell has a characteristic timescale of $\sim$200~ps, followed by an ion tail lasting about 50~ns~\cite{urwell-timing}. The thick amplification gap (0.4 mm) of the RPWELL leads to a much longer signal formation: electron component of $\approx$10~ns followed by an ion tail that can extend to the microsecond range~\cite{Bhattacharya:2018sqx}.

\begin{figure}
    \centering
    \begin{subfigure}[b]{0.32\textwidth}
    \includegraphics[width=\textwidth]{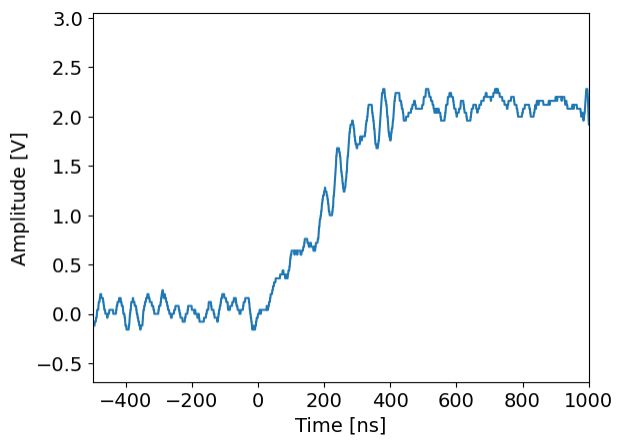}
    \caption{Micromegas}
    \label{fig:signal-mm}
    \end{subfigure}
    \begin{subfigure}[b]{0.32\textwidth}
    \includegraphics[width=\textwidth]{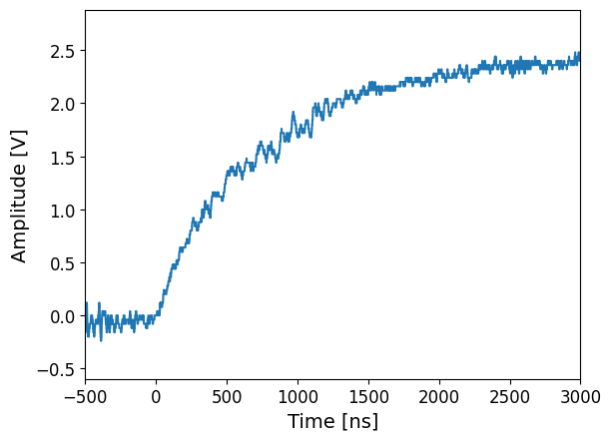}
    \caption{RPWELL}
    \label{fig:signal-rpwell}
    \end{subfigure}
    \begin{subfigure}[b]{0.32\textwidth}
    \includegraphics[width=\textwidth]{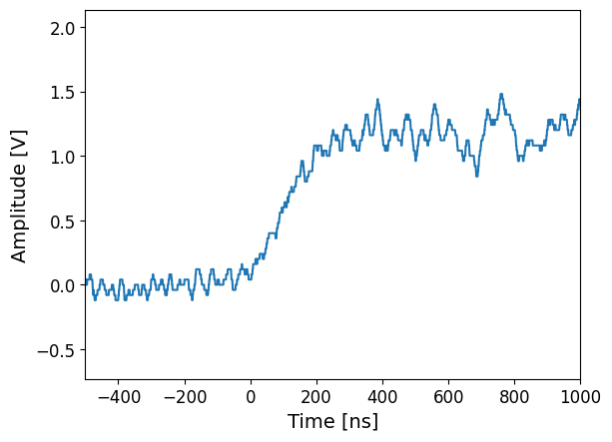}
    \caption{\urwell}
    \label{fig:signal-urwell}
    \end{subfigure}
    \caption{Representative signal shapes recorded from Micromegas, $\mu$-RWELL, and RPWELL detectors using a charge-sensitive preamplifier.}
    \label{fig:signals}
\end{figure}

Figure~\ref{fig:signals} shows typical cosmic muon-induced signals in the Micromegas (\ref{fig:signal-mm}), RPWELL (\ref{fig:signal-rpwell}) and \urwell (\ref{fig:signal-urwell}). The signals were recorded using charge-sensitive preamplifier\footnote{Cremat CR-110-R2.1 model} (CSP) and read out with an oscilloscope.

The RPWELL signal is clearly an order of magnitude longer in duration. As a result, when read out with the VMM3a ASIC (with 200 ns integration), only $30\%$  of the total charge is integrated.

\bibliographystyle{JHEP}
\bibliography{refs}

\providecommand{\href}[2]{#2}\begingroup\raggedright\begin{thebibliography}{10}

\bibitem{Alviggi:2024lir}
{M. Alviggi et al
  }\href{https://doi.org/10.1088/1748-0221/20/01/P01012}{\emph{JINST}
  {\bfseries 20} (2025) P01012}
  [\href{https://arxiv.org/abs/2411.17202}{{\ttfamily 2411.17202}}].

\bibitem{Rubin:2013jna}
{A. Rubin et al
  }\href{https://doi.org/10.1088/1748-0221/8/11/P11004}{\emph{JINST} {\bfseries
  8} (2013) P11004} [\href{https://arxiv.org/abs/1308.6152}{{\ttfamily
  1308.6152}}].

\bibitem{Bencivenni:2024ryx}
{G. Bencivenni et al
  }\href{https://doi.org/10.1088/1748-0221/19/02/C02057}{\emph{JINST}
  {\bfseries 19} (2024) C02057}.

\bibitem{jash2024discharge}
{A. Jash et al }\href{https://doi.org/10.1016/j.nima.2024.169319}{\emph{Nucl.
  Instrum. Meth. A} {\bfseries 1064} (2024) }.

\bibitem{Scharenberg:2024abh}
{L. Scharenberg et al
  }\href{https://doi.org/10.1088/1748-0221/19/05/P05053}{\emph{JINST}
  {\bfseries 19} (2024) P05053}
  [\href{https://arxiv.org/abs/2402.03899}{{\ttfamily 2402.03899}}].

\bibitem{MuonColliderendorsedbytheInternational:2025zcf}
{A. Stamerra et al
  }\href{https://doi.org/10.1088/1748-0221/20/06/C06019}{\emph{JINST}
  {\bfseries 20} (2025) C06019}.

\bibitem{Zavazieva:2025rsj}
{D. Zavazieva et al
  }\href{https://doi.org/10.1088/1748-0221/20/03/C03026}{\emph{JINST}
  {\bfseries 20} (2025) C03026}.

\bibitem{Longo:2024ntu}
{L. Longo et al }\href{https://doi.org/10.22323/1.476.1082}{\emph{PoS}
  {\bfseries ICHEP2024} (2025) 1082}.

\bibitem{WANG2010151}
{Y. Wang et al
  }\href{https://doi.org/https://doi.org/10.1016/j.nima.2010.04.056}{\emph{{Nucl.
  Instrum. Meth. A}} {\bfseries 621} (2010) 151}.

\bibitem{zavazieva2023towards}
{D. Zavazieva et al
  }\href{https://doi.org/10.1088/1748-0221/18/08/P08009}{\emph{JINST}
  {\bfseries 18} (2023) P08009}.

\bibitem{Altunbas:2002ds}
{C. Altunbas et al
  }\href{https://doi.org/10.1016/S0168-9002(02)00910-5}{\emph{Nucl. Instrum.
  Meth. A} {\bfseries 490} (2002) 177}.

\bibitem{scharenberg2022development}
{L. Scharenberg et al
  }\href{https://doi.org/10.1088/1748-0221/17/12/C12014}{\emph{JINST}
  {\bfseries 17} (2022) C12014}.

\bibitem{vmmsdat}
{D. Pfeiffer et al}.
\newblock \href{https://github.com/ess-dmsc/vmm-sdat}{GitHub "vmm-sdat –
  VMM3a/SRS Data Analysis Tool" 2024}.

\bibitem{tb24-tracking}
{E.B. Shields and L. Moleri}.
\newblock \href{https://gitlab.cern.ch/eshields/tb24-tracking}{GitLab
  "TB24-tracking" 2024}.

\bibitem{corrympgd}
{M. Borysova}.
\newblock \href{https://gitlab.cern.ch/maboryso/corryvreckan-mpgd.git}{GitLab
  "CorryvreckanMPGD" 2024}.

\bibitem{dannheim2021corryvreckan}
{D. Dannheim et al
  }\href{https://doi.org/10.1088/1748-0221/16/03/P03008}{\emph{JINST}
  {\bfseries 16} (2021) P03008}.

\bibitem{ALEXEEV201796}
{Levorato, S. et al
  }\href{https://doi.org/https://doi.org/10.1016/j.nima.2017.02.013}{\emph{Nucl.
  Instrum. Meth. A} {\bfseries 876} (2017) 96}.

\bibitem{Bortfeldt:2010development}
J.~Bortfeldt.
\newblock
  \href{https://www-static.etp.physik.uni-muenchen.de/dokumente/thesis/dipl_bortfeldt.pdf}{Thesis
  at Ludwig-Maximilians-Universitat Munchen 2010}.

\bibitem{urwell-timing}
{M. Poli Lener et al}, \emph{{The micro-RWELL}},  2018.
\newblock \url{https://agenda.infn.it/event/14816/}.

\bibitem{Bhattacharya:2018sqx}
{P. Bhattacharya et al
  }\href{https://doi.org/10.1016/j.nima.2018.10.214}{\emph{Nucl. Instrum. Meth.
  A} {\bfseries 916} (2019) 125}
  [\href{https://arxiv.org/abs/1810.12597}{{\ttfamily 1810.12597}}].

\end{thebibliography}\endgroup

\end{document}